\documentclass[12pt]{article} \usepackage{epsfig,amsfonts,amssymb}
 
 \newcommand{\be}{\begin{equation}}
  \newcommand{\ee}{\end{equation}} \newcommand{\bea}{\begin{eqnarray}}
  \newcommand{\eea}{\end{eqnarray}}
\newcommand{\beas}{\begin{eqnarray*}}
  \newcommand{\eeas}{\end{eqnarray*}}

\def\scriptlap{{\kern1pt\vbox{\hrule height 0.8pt\hbox{\vrule width
        0.8pt \hskip2pt\vbox{\vskip 4pt}\hskip 2pt\vrule width
        0.4pt}\hrule height 0.4pt} \kern1pt}} 
  
\def\kkappa{{\beta}} 
 \def\Biggg#1{{\hbox{$\left#1\vbox to
        25pt{}\right.\n@space$}}} \def\n@space{\nulldelimiterspace=0pt
  \m@th} \def\m@th{\mathsurround = 0pt}
\newcommand{\ta}{\tilde{\alpha}}
 \newcommand{\PP}{Q_2}
\newcommand{\Q}{Q_1} \newcommand{\aod}{\frac{\alpha}{\alpha-\ta}}
\newcommand{\nosum}{\qquad\mbox{(no summation)}}
\newcommand{\ha}{\frac{1}{2}}
\newcommand{\textha}{{\textstyle{\frac{1}{2}}}}

\begin{document}

\voffset -0.7 true cm \hoffset 1.1 true cm \topmargin 0.0in
\evensidemargin 0.0in \oddsidemargin 0.0in \textheight 8.6in
\textwidth 7.25in \parskip 10 pt

\begin{titlepage}
  \begin{flushright}
    {\small TIFR/TH/05-25} \\
    {\small hep-th/0507096}
  \end{flushright}

  \begin{center}

    \vspace{26mm}

    {\LARGE \bf Non-Supersymmetric Attractors}

    \vspace{10mm}

    Kevin Goldstein, Norihiro Iizuka, Rudra P. Jena and Sandip P.
    Trivedi

    \vspace{5mm}

    {\small \sl Tata Institute of Fundamental Research} \\
    {\small \sl Homi Bhabha Road, Mumbai, 400 005, INDIA} \\
    {\small \tt kevin, iizuka, rpjena, sandip@theory.tifr.res.in}
    \vspace{10mm}

  \end{center}

  \vskip 0.3 cm \centerline{\bf Abstract} \vspace{5mm}
  \noindent
  We consider theories with gravity, gauge fields and scalars in
  four-dimensional asymptotically flat space-time.  By studying the
  equations of motion directly we show that the attractor mechanism
  can work for non-supersymmetric extremal black holes.  Two
  conditions are sufficient for this, they are conveniently stated in
  terms of an effective potential involving the scalars and the
  charges carried by the black hole.  Our analysis applies to black
  holes in theories with ${\cal N}\leqslant 1$ supersymmetry, as well as
  non-supersymmetric black holes in theories with ${\cal N}=2$
  supersymmetry. Similar results are also obtained for extremal black
  holes in asymptotically Anti-de Sitter space and in higher
  dimensions.
 
\end{titlepage}
\setcounter{tocdepth}{2}
\tableofcontents
\section{Introduction}

Black holes in ${\cal N}=2$ supersymmetric theories are known to
exhibit a fascinating phenomenon called the the attractor mechanism.
There is a family of black hole solutions in these theories which are
spherically symmetric, extremal black holes, with double-zero horizons
\footnote{By a double-zero horizon we mean a horizon for which the
  surface gravity vanishes because the $g_{00}$ component of the
  metric has a double-zero (in appropriate coordinates), as in an
  extremal Reissner Nordstrom black hole.}.  In these solutions
several moduli fields are drawn to fixed values at the horizon of the
black hole regardless of the values they take at asymptotic infinity.
The fixed values are determined entirely by the charges carried by the
black hole.  This phenomenon was first discussed by
\cite{Ferrara:1995ih} and has been studied quite extensively since
then
\cite{Cvetic:1995bj,Strominger:1996kf,Ferrara:1996dd,Ferrara:1996um,Cvetic:1996zq,Ferrara:1997tw,Gibbons:1996af,Denefa,Denef:2001xn}.
It has gained considerable attention recently due to the conjecture of
\cite{Ooguri:2004zv} and related developments
\cite{LopesCardoso:1998wt,Dabholkar:2004yr,Ooguri:2005vr,Dijkgraaf:2005bp}.

So far the attractor phenomenon has been studied almost exclusively in
the context of BPS black holes in the ${\cal N}=2$ theories.  The aim
of this paper is to examine if it is more general and can happen for
non-supersymmetric black holes as well. These black holes might be
solutions in theories which have no supersymmetry or might be
non-supersymmetric solutions in ${\cal N} \geqslant 1$ supersymmetric
theories.

There are two motivations for this investigation.  First, a
non-supersymmetric attractor mechanism might help in the study of
non-supersymmetric black holes, especially their entropy. Second,
given interesting parallels between flux compactifications and the
attractor mechanisms, a non-supersymmetric attractor phenomenon might
lead to useful lessons for non-supersymmetric flux compactifications.
For example, it could help in finding dual descriptions of such
compactifications. This might help to single out vacua with a small
cosmological constant. Or it might suggest ways to weight vacua with
small cosmological constants preferentially while summing over all of
them \footnote{For a recent attempt along these lines where
  supersymmetric compactifications have been considered, see
  \cite{Ooguri:2005vr,Dijkgraaf:2005bp}.}.  These lessons would be
helpful in light of the vast number of vacua that have been recently
uncovered in string theory \cite{KKLT}.

An intuitive argument for the attractor mechanism is as follows. One
expects that the total number of microstates corresponding to an
extremal black hole is determined by the quantised charges it carries,
and therefore does not vary continuously. If the counting of
microstates agrees with the Bekenstein-Hawking entropy, that is the
horizon area, it too should be determined by the charges alone.  This
suggests that the moduli fields which determine the horizon area take
fixed values at the horizon, and these fixed values depend only on the
charges, independent of the asymptotic values for the moduli.  While
this argument is only suggestive what is notable for the present
discussion is that it does not rely on supersymmetry. This provides
further motivation to search for a non-supersymmetric version of the
attractor mechanism.

The theories we consider in this paper consist of gravity, gauge
fields and scalar fields.  The scalars determine the gauge couplings
and there by couple to the gauge fields.  It is important that the
scalars do not have a potential of their own that gives them in
particular a mass.  Such a potential would mean that the scalars are
no longer moduli.
 
We first study black holes in asymptotically flat four dimensions.
Our main result is to show that the attractor mechanism works quite
generally in such theories provided two conditions are met. These
conditions are succinctly stated in terms of an ``effective
potential'' $V_{eff}$ for the scalar fields, $\phi_i$.  The effective
potential is proportional to the energy density in the electromagnetic
field and arises after solving for the gauge fields in terms of the
charges carried by the black hole, as we explain in more detail below.
The two conditions that need to be met are the following. First, as a
function of the moduli fields $V_{eff}$ must have a critical point,
$\partial_iV_{eff}(\phi_{i0})=0$. And second, the matrix of second
derivatives of the effective potential at the critical point,
$\partial_{ij}V_{eff}(\phi_{i0})$, must be have only positive
eigenvalues. The resulting attractor values for the moduli are the
critical values, $\phi_{i0}$. And the entropy of the black hole is
proportional to $V(\phi_{i0})$, and is thus independent of the
asymptotic values for the moduli. It is worth noting that the two
conditions stated above are met by BPS black hole attractors in an
${\cal N}=2$ theory.

The analysis for BPS attractors simplifies greatly due to the use of
the first order equations of motion.  In the non-supersymmetric
context one has to work with the second order equations directly and
this complicates the analysis.  We find evidence for the attractor
mechanism in three different ways.  First, we analyse the equations
using perturbation theory.  The starting point is a black hole
solution, where the asymptotic values for the moduli equal their
critical values.  This gives rise to an extremal Reissner Nordstrom
black hole. By varying the asymptotic values a little at infinity one
can now study the resulting equations in perturbation theory. Even
though the equations are second order, in perturbation theory they are
linear, and this makes them tractable.  The analysis can be carried
out quite generally for any effective potential for the scalars and
shows that the two conditions stated above are sufficient for the
attractor phenomenon to hold.
 
Second, we carry out numerical analysis.  This requires a specific
form of the effective potential, but allows us to go beyond the
perturbative regime.  The numerical analysis corroborates the
perturbation theory results mentioned above.  In simple cases we have
explored so far, we have found evidence for a only single basin of
attraction, although multiple basins must exist in general as is
already known from the SUSY cases.

Finally, in some special cases, we solve the equations of motion
exactly by mapping them a solvable Toda system. This allows us to
study the black hole solutions in these special cases in some depth.
Once again, in all the cases we have studied, we can establish the
attractor phenomenon.

It is straightforward to generalise these results to other settings.
We find that the attractor phenomenon continues to hold in Anti-de
Sitter space (AdS) and also in higher dimensions, as long as the two
conditions mentioned above are valid for a suitable defined effective
potential.  There is also possibly an attractor mechanism in de Sitter
space (dS), but in the simplest of situations analysed here some
additional caveats have to be introduced to deal with infrared
divergences in the far past (or future) of dS space.

This paper is structured as follows. Black holes in asymptotically
flat four dimensional space are analysed first, in sections 2,3,4.
The discussion is extended to asymptotically flat space-times of
higher dimension in section 5.  Asymptotically AdS space is discussed
next in section 6.

As was mentioned above our analysis in the asymptotically flat and AdS
cases is based on theories which have no potential for the scalars so
that their values can vary at infinity.  Some comments on this are
contained in Section 7.  With ${\cal N}\geqslant 1$ SUSY such theories can
arise, with the required couplings between scalars and gauge fields,
and are at least technically natural. In the absence of supersymmetry
there is no natural way to arrange this and our study is more in the
nature of a mathematical investigation.  We follow in section 8, with
some comments on the attractor phenomenon in dS.  Finally, in section
9 we show that non-extremal black holes do not have an attractor
mechanism.  Thus, the double-zero nature of the horizon is essential
to draw the moduli to fixed values.

Several important intermediate steps in the analysis are discussed in
appendices \ref{sec:appa}-\ref{sec:ads}.

Some important questions are left for the future. First, we have not
analysed the stability of these black hole solutions.  It is unlikely
that there are any instabilities at least in the S-wave sector. We do
not attempt a general analysis of small fluctuations here.  Second, in this paper  we
have not analysed string theory situations where such
non-supersymmetric black holes can arise \cite{TT3}.  This could include both
critical and non-critical string theory. In case of ${\cal
  N}=1$ supersymmetry it would be interesting to explore if there is
partial restoration of supersymmetry at the horizon. Given the
rotational invariance of the solutions one can see that no
supersymmetry is preserved in-between asymptotic infinity and the
horizon in this case. 

Let us also briefly comment on some of the literature of especial
relevance. The importance of the effective potential, $V_{eff}$, for
${\cal N}=2$ black holes was emphasised in \cite{Denefa}. Some
comments pertaining to the non-supersymmetric case can be found for
example in \cite{Ferrara:1997tw}. A similar  analysis using an effective  one dimensional theory,
and the Gauss-Bonett term, was carried out in \cite{Sen:2005kj}. Finally, while the thrust of the
analysis is different, our results are quite closely related to those
in \cite{Sen:2005wa} which appeared while this paper was in
preparation (see also \cite{Kraus:2005vz} for the 3-dimensional case).
In \cite{Sen:2005wa} the entropy (including higher derivative
corrections) is obtained from the gauge field Lagrangian after
carrying out a Legendre transformation with respect to the electric
parameters. This is similar to our result which is based on
$V_{eff}$. As was mentioned above, $V_{eff}$, is proportional to the
electro-magnetic energy density i.e., the Hamiltonian density of the
electro-magnetic fields, and is derived from the Lagrangian by doing a
canonical transformation with respect to the gauge fields.
For an action   with only  two-derivative terms, our  results
and those  in \cite{Sen:2005wa} agree \cite{IJ}.

\section{Attractor in Four-Dimensional Asymptotically Flat Space}

\subsection{Equations of Motion }

In this section we consider gravity in four dimensions with $U(1)$
gauge fields and scalars. The scalars are coupled to gauge fields with
dilaton-like couplings.  It is important for the discussion below that
the scalars do not have a potential so that there is a moduli space
obtained by varying their values.

The action we start with has the form,
\begin{equation}
  S=\frac{1}{\kappa^{2}}\int d^{4}x\sqrt{-G}(R-2(\partial\phi_i)^{2}-
  f_{ab}(\phi_i)F^a_{\mu \nu} F^{b \ \mu \nu})
\end{equation}
Here the index $i$ denotes the different scalars and $a,b$ the
different gauge fields and $F^a_{\mu \nu}$ stands for the field
strength of the gauge field. $f_{ab}(\phi_i)$ determines the gauge
couplings, we can take it to be symmetric in $a,b$ without loss of
generality.

The Lagrangian is
\begin{equation}
  {\mathcal{L}}=(R-2(\partial\phi_i)^{2}-f_{ab}(\phi_i) F^a_{\mu \nu} F^{b \ \mu \nu})
\end{equation}
Varying the metric gives \footnote{In our notation $G_{\mu\nu}$ refers to the components of the metric.},
\begin{equation}
  R_{\mu\nu}-2\partial_{\mu}\phi_i\partial_{\nu}\phi_i
  =2 f_{ab}(\phi_i) F^a_{\phantom{a}\mu\lambda} F^{b\phantom{\nu}\lambda}_{\phantom{b}\nu}
  +\frac{1}{2} G_{\mu\nu}\mathcal{L}
  \label{eq:metric1}
\end{equation}
The trace of the above equation implies
\begin{equation}
  R-2(\partial\phi_i)^{2}=0\label{trace}
\end{equation}
The equations of motion corresponding to the metric, dilaton and the
gauge fields are then given by,
\begin{eqnarray}
  R_{\mu\nu}-2\partial_{\mu}\phi_i\partial_{\nu}\phi_i 
  & = &   f_{ab}(\phi_i) \left(2F^a_{\phantom{a}\mu\lambda}F^{b\phantom{\nu}\lambda}_{\phantom{b}\nu}-
    {\textstyle \frac{1}{2}}G_{\mu\nu}F^a_{\phantom{a}\kappa\lambda} F^{b \kappa\lambda} \right) \label{motion} \\
  \frac{1}{\sqrt{-G}}\partial_{\mu}(\sqrt{-G}\partial^{\mu}\phi_i) 
  & = & {1 \over 4} \partial_i(f_{ab}) F^a _{\phantom{a}\mu\nu} F^{b \mu\nu} \label{dilatonmotion} \\
  \partial_{\mu}(\sqrt{-G}f_{ab}(\phi_i) F^{b \mu\nu}) & = & 0.\nonumber 
\end{eqnarray}
The Bianchi identity for the gauge field is,
\begin{equation}
\label{bi}
\partial_\mu F_{\nu \rho} + \partial_\nu F_{\rho \mu} + \partial_\rho F_{\mu \nu}=0.
\end{equation}
We now assume all quantities to be function of $r$.  To begin, let us
also consider the case where the gauge fields have only magnetic
charge, generalisations to both electrically and magnetically charged
cases will be discussed shortly.  The metric and gauge fields can then
be written as,
\begin{eqnarray}
  ds^{2} & = & -a(r)^{2}dt^{2}+a(r)^{-2}dr^{2}+b(r)^{2}d\Omega^{2}\label{metric2}\\
  F^a & = & Q_m^{a}sin\theta d\theta\wedge d\phi \label{fstrength}\end{eqnarray}
Using the equations of motion we then get, 
\begin{eqnarray}
  R_{tt} & = & \frac{a^{2}}{b^{4}} V_{eff}(\phi_i) \label{energyd} \\
  R_{\theta\theta} & = & \frac{1}{b^{2}} V_{eff}(\phi_i) \label{stressd} 
\end{eqnarray}
where,
\begin{equation}
  \label{defpot}
  V_{eff}(\phi_i) \equiv f_{ab}(\phi_i) Q_m^a Q_m^b.
\end{equation}
This function, $V_{eff}$, will play an important role in the
subsequent discussion.  We see from eq.(\ref{energyd}) that up to an
overall factor it is the energy density in the electromagnetic field.
Note that $V_{eff}(\phi_i)$ is actually a function of both the scalars
and the charges carried by the black hole.

The relation, $R_{tt}=\frac{a^{2}}{b^{2}}R_{\theta\theta}$, after
substituting the metric ansatz implies that,
\begin{equation}
  \label{eq1}
  (a^{2}(r)b^{2}(r))^{''}=2.
\end{equation}

The $R_{rr}-{G^{tt}\over G^{rr}} R_{tt}$ component of the Einstein
equation gives
\begin{equation}
  \frac{b^{''}}{b}=-(\partial_{r}\phi)^{2}. \label{eq2}
\end{equation}
Also the $R_{rr}$ component itself yields a first order ``energy''
constraint,
\begin{equation}
  -1+a^{2}b^{'2}+\frac{a^{2'}b^{2'}}{2}=\frac{-1}{b^{2}}(V_{eff}(\phi_i))+a^{2}b^{2}
  (\phi')^{2}\label{constraint}
\end{equation}

Finally, the equation of motion for the scalar $\phi_i$ takes the
form,
\begin{equation}
  \label{eomdil}
  \partial_{r}(a^{2}b^{2}\partial_{r}\phi_i)=\frac{\partial_iV_{eff} }{2b^{2}}.
\end{equation}
We see that $V_{eff}(\phi_i)$ plays the role of an ``effective
potential '' for the scalar fields.

Let us now comment on the case of both electric and magnetic charges.
In this case one should also include ``axion'' type couplings and the
action takes the form,
\begin{equation}
  S=\frac{1}{\kappa^{2}}\int d^{4}x\sqrt{-G}(R-2(\partial\phi_i)^{2}-
  f_{ab}(\phi_i)F^a_{\phantom{a}\mu \nu} F^{b \, \mu \nu} 
  -{\textstyle{1 \over 2}} {\tilde f}_{ab}(\phi_i) F^a_{\phantom{a}\mu \nu} 
  F^b_{\phantom{b}\rho \sigma} \epsilon^{\mu \nu \rho \sigma} ).
  \label{actiongen}
\end{equation}
We note that ${\tilde f}_{ab}(\phi_i)$ is a function independent of
$f_{ab}(\phi_i)$, it can also be taken to be symmetric in $a,b$
without loss of generality.

The equation of motion for the metric which follows from  this action
is unchanged from eq.(\ref{motion}). While the equations of motion for the dilaton and the gauge field now
 take the form,
\begin{equation}
\label{genedg}
\frac{1}{\sqrt{-G}}\partial_{\mu}(\sqrt{-G}\partial^{\mu}\phi_i)
   =  \textstyle{1 \over 4} \partial_i(f_{ab}) F^a _{\phantom{a}\mu\nu} F^{b\, \mu\nu} +\textstyle{1\over 8} 
   \partial_i({\tilde f}_{ab}) F^a_{\phantom{a}\mu\nu} F^b_{\phantom{b}\rho \sigma} \epsilon^{\mu\nu\rho\sigma} \label{gendilatonmotion}
\end{equation}
\begin{equation}
  \partial_{\mu}\left(\sqrt{-G}\left(f_{ab}(\phi_i) F^{b\, \mu\nu} 
+ \textstyle{1\over 2} {\tilde f}_{ab}F^b_{\phantom{b}\rho\sigma}
\epsilon^{\mu\nu\rho\sigma}\right) \right)  =  0.\label{geneomgf}
\end{equation}
 
With both electric and magnetic charges the gauge fields take the
form,
\begin{equation}
  \label{fstrenghtgen}
  F^a=f^{ab}(\phi_i)(Q_{eb}-{\tilde f}_{bc}Q^c_m) {1\over b^2} dt\wedge dr + Q_m^a sin \theta  d\theta \wedge d\phi, 
\end{equation} 
where $Q_m^a, Q_{ea}$ are constants that determine the magnetic and
electric charges carried by the gauge field $F^a$, and $f^{ab}$ is the
inverse of $f_{ab}$ \footnote{We assume that $f_{ab}$ is invertible.
  Since it is symmetric it is always diagonalisable. Zero eigenvalues
  correspond to gauge fields with vanishing kinetic energy terms,
  these can be omitted from the Lagrangian.}.  
It is easy to see that
this solves the Bianchi identity eq.(\ref{bi}), and the equation of motion for the
gauge fields eq.(\ref{geneomgf}).

A little straightforward algebra shows that the Einstein equations for
the metric and the equations of motion for the scalars take the same
form as before, eq.(\ref{eq1}, \ref{eq2}, \ref{constraint},
\ref{eomdil}), with $V_{eff}$ now being given by,
\begin{equation}
  \label{defpotgen}
  V_{eff}(\phi_i)=f^{ab}(Q_{ea}-{\tilde f}_{ac}Q^c_m)(Q_{eb}- {\tilde f}_{bd}Q^d_m)+f_{ab}Q^a_mQ^b_m.
\end{equation}
As was already noted in the special case of only magnetic charges,
$V_{eff}$ is proportional to the energy density in the electromagnetic
field and therefore has an immediate physical significance.  It is
invariant under duality transformations which transform the electric
and magnetic fields to one-another.

Our discussion below will use (\ref{eq1}, \ref{eq2}, \ref{constraint},
\ref{eomdil}) and will apply to the general case of a black hole
carrying both electric and magnetic charges.

It is also worth mentioning that the equations of motion,
eq.(\ref{eq1}, \ref{eq2}, \ref{eomdil}) above can be derived from a
one-dimensional action,
\begin{eqnarray}
  S=\frac{2}{\kappa^{2}}\int dr\left((a^{2}b)^{'} b^{'}
    -a^{2}b^{2}(\phi')^{2}-\frac{ V_{eff}(\phi_i)}{b^{2}}\right).\label{action}
\end{eqnarray}
The constraint, eq.(\ref{constraint}) must be imposed in addition.

One final comment before we proceed. The eq.(\ref{actiongen}) can be
further generalised to include non-trivial kinetic energy terms for
the scalars of the form,
\begin{equation}
  \label{compkinetic}
  \int d^4x \sqrt{-G}\left(- g_{ij}(\phi_k) \partial\phi^i \partial\phi^j\right).
\end{equation}
The resulting equations are easily determined from the discussion
above by now contracting the scalar derivative terms with the metric
$g_{ij}$.  The two conditions we obtain in the next section for the
existence of an attractor are not altered due to these more general
kinetic energy terms.

\subsection{Conditions for an Attractor \label{sec:cond:attr}}
We can now state the two conditions which are sufficient for the
existence of an attractor.  First, the charges should be such that the
resulting effective potential, $V_{eff}$, given by
eq.(\ref{defpotgen}), has a critical point.  We denote the critical
values for the scalars as $\phi_i=\phi_{i0}$.  So that,
\begin{equation}
  \label{critical}
  \partial_iV_{eff}(\phi_{i0})=0
\end{equation}
Second, the matrix of second derivatives of the potential at the
critical point,
\begin{equation}
  \label{massmatrix}
  M_{ij}={1\over 2} \partial_i\partial_jV_{eff}(\phi_{i0})
\end{equation}
should have positive eigenvalues. Schematically we write,
\begin{equation}
  \label{positive}
  M_{ij}>0
\end{equation}
Once these two conditions hold, we show below that the attractor
phenomenon results.  The attractor values for the scalars are
\footnote{Scalars which do not enter in $V_{eff}$ are not fixed by the
  requirement eq.(\ref{critical}). The entropy of the extremal black
  hole is also independent of these scalars.}  $\phi_i=\phi_{i0}$.

The resulting horizon radius is given by,
\begin{equation}
  \label{RH}
  b_H^2=V_{eff}(\phi_{i0})
\end{equation}
and the entropy is
 \be
\label{BH}
S_{BH}={1 \over 4} A = \pi b_H^2. 
\ee

There is one special solution which plays an important role in the
discussion below.  From eq.(\ref{eomdil}) we see that one can
consistently set $\phi_i=\phi_{i0}$ for all values of $r$. The
resulting solution is an extremal Reissner Nordstrom (ERN) Black hole.
It has a double-zero horizon. In this solution $\partial_r\phi_{i}=0$,
and $a,b$ are
\begin{eqnarray}
  a_0(r)=\left(1-\frac{r_{H}}{r}\right),\,\,\
  b_0(r)=r \label{ern} \end{eqnarray}
where $r_{H}$ is the horizon radius.
We see that $a_0^2, (a_0^2)'$ vanish at the horizon while $b_0, b_0'$ are finite there. 
From eq.(\ref{constraint})  it follows then that the horizon radius $b_H$ is indeed given by
\begin{equation}
  \label{horizon}
  r_H^2=b_H^2= V_{eff}(\phi_{i0}),
\end{equation}
and the black hole entropy is eq.(\ref{BH}).
 
If the scalar fields take values at asymptotic infinity which are
small deviations from their attractor values we show below that a
double-zero horizon black hole solution continues to exist. In this
solution the scalars take the attractor values at the horizon, and
$a^2, (a^2)'$ vanish while $b, b'$ continue to be finite there.  From
eq.(\ref{constraint}) it then follows that for this whole family of
solutions the entropy is given by eq.(\ref{BH}) and in particular is
independent of the asymptotic values of the scalars.

For simple potentials $V_{eff}$ we find only one critical point. In
more complicated cases there can be multiple critical points which are
attractors, each of these has a basin of attraction.

One comment is worth making before moving on.  A simple example of a
system which exhibits the attractor behaviour consists of one scalar
field $\phi$ coupled to two gauge fields with field strengths, $F^a,
a= 1, 2$. The scalar couples to the gauge fields with dilaton-like
couplings,
\begin{equation}
  \label{se}
  f_{ab}(\phi)=e^{\alpha_a \phi} \delta_{ab}.
\end{equation}
If only magnetic charges are turned on,
\begin{equation}
  \label{seffpot}
  V_{eff}=e^{\alpha_1 \phi} (Q_1)^2 + e^{\alpha_2 \phi} (Q_2)^2.
\end{equation}
(We have suppressed the subscript ``$m$'' on the charges).  For a
critical point to exist $\alpha_1$ and $\alpha_2$ must have opposite
sign.  The resulting critical value of $\phi$ is given by,
\begin{equation}
  e^{\phi_{0}}=\left(-\frac{\alpha_2 (Q_{2})^{2}}{\alpha_1 (Q_{1})^{2}}\right)^{\frac{1}{\alpha_1-\alpha_2}}
  \label{attractorexample}
\end{equation}
The second derivative, eq.(\ref{massmatrix}) now is given by
\begin{equation}
  \label{sdexample}
  {\partial^2V_{eff}\over \partial \phi^2 }=-2 \alpha^1 \alpha^2
\end{equation}
and is positive if $\alpha_1,\alpha_2$ have opposite sign.

This example will be useful for studying the behaviour of perturbation
theory to higher orders and in the subsequent numerical analysis.

As we will discuss further in section 7, a Lagrangian with dilaton-like couplings of the type in eq.(\ref{se}),
and additional axionic terms ( which can be consistently set to zero if only magnetic charges are turned on),
 can always  be embedded in a theory with ${\cal N}=1$ supersymmetry.
But for generic values of $\alpha$ we do not expect to be able to embed it in an ${\cal N}=2$ theory.
The resulting extremal black hole, for generic $\alpha$, will also then not  be a  BPS state.

\subsection{Comparison with the ${\cal N}=2$ Case}
It is useful to compare the discussion above with the special case of
a BPS black hole in an ${\cal N}=2$ theory.  The role of the effective
potential, $V_{eff}$ for this case was emphasised by Denef,
\cite{Denefa}.  It can be expressed in terms of a superpotential $W$
and a Kahler potential $K$ as follows:
\begin{equation}
  \label{comparison}
  V_{eff}=e^K[K^{i \bar j}D_i W (D_j W)^* + |W|^2],
\end{equation}
where $D_iW\equiv \partial_iW+\partial_iK W$.  The attractor equations
take the form,
\begin{equation}
  \label{attractorsusy}
  D_iW=0
\end{equation}
And the resulting entropy is given by
\begin{equation}
  \label{susyentropy}
  S_{BH}=\pi |W|^2 e^K.
\end{equation}
with the superpotential evaluated at the attractor values.

It is easy to see that if eq.(\ref{attractorsusy}) is met then the
potential is also at a critical point, $\partial_i V_{eff}=0$. A
little more work also shows that all eigenvalues of the second
derivative matrix, eq.(\ref{massmatrix}) are also positive in this
case. Thus the BPS attractor meets the two conditions mentioned above.
We also note that from eq.(\ref{comparison}) the value of $V_{eff}$ at
the attractor point is $V_{eff}=e^K|W|^2$.  The resulting black hole
entropy eq.(\ref{RH}, \ref{BH}) then agrees with
eq.(\ref{susyentropy}).

We now turn to a more detailed analysis of the attractor conditions
below.

\subsection{Perturbative Analysis}
\subsubsection{A Summary}

The essential idea in the perturbative analysis is to start with the
extremal RN black hole solution described above, obtained by setting
the asymptotic values of the scalars equal to their critical values,
and then examine what happens when the scalars take values at
asymptotic infinity which are somewhat different from their attractor
values, $\phi_{i}=\phi_{i0}$.

We first study the scalar field equations to first order in the
perturbation, in the ERN geometry without including backreaction.  Let
$\phi_i$ be a eigenmode of the second derivative matrix
eq.(\ref{massmatrix}) \footnote{More generally if the kinetic energy
  terms are more complicated, eq.(\ref{compkinetic}), these eigenmodes
  are obtained as follows. First, one uses the metric at the attractor
  point, $g_{ij}(\phi_{i0})$, and calculates the kinetic energy terms.
  Then by diagonalising and rescaling one obtains a basis of
  canonically normalised scalars.  The second derivatives of $V_{eff}$
  are calculated in this basis and gives rise to a symmetric matrix,
  eq.(\ref{massmatrix}). This is then diagonalised by an orthogonal
  transformation that keeps the kinetic energy terms in canonical
  form. The resulting eigenmodes are the ones of relevance here.}.
Then denoting, $\delta \phi_i \equiv \phi_i -\phi_{i0}$, neglecting
the gravitational backreaction, and working to first order in $\delta
\phi_i$, we find that eq.(\ref{eomdil}) takes the form, 
\be
\label{nhpert}
\partial_r\left((r-r_H)^2\partial_r (\delta \phi_i)\right) = {\beta_i^2 \delta \phi_i
  \over r^2 }, 
\ee 
where $\beta_i^2 $ is the relevant eigenvalue of
${1 \over 2} \partial_i\partial_j V(\phi_{i0})$.  In the vicinity of
the horizon, we can replace the factor $1/r^2$ on the r.h.s by a
constant and as we will see below, eq.(\ref{nhpert}), has one solution
that is well behaved and vanishes at the horizon provided $\beta_i^2
\geqslant 0$.  Asymptotically, as $r\rightarrow \infty$, the effects of the
gauge fields die away and eq.(\ref{nhpert}) reduces to that of a free
field in flat space. This has two expected solutions, $\delta \phi_i
\sim constant$, and $\delta \phi_i \sim 1/r$, both of which are well
behaved.  It is also easy to see that the second order differential
equation is regular at all points in between the horizon and infinity.
So once we choose the non-singular solution in the vicinity of the
horizon it can be continued to infinity without blowing up.

Next, we include the gravitational backreaction. The first order
perturbations in the scalars source a second order change in the
metric.  The resulting equations for metric perturbations are regular
between the horizon and infinity and the analysis near the horizon and
at infinity shows that a double-zero horizon black hole solution
continues to exist which is asymptotically flat after including the
perturbations.

In short the two conditions, eq.(\ref{critical}), eq.(\ref{positive}),
are enough to establish the attractor phenomenon to first non-trivial
order in perturbation theory.  

In  4-dimensions, for an  effective potential which can be expanded 
in a power series about its  minimum, one can in principle solve for the perturbations 
analytically to all orders in perturbation theory.  We illustrate this below for the simple case 
of dilaton-like couplings, eq.(\ref{se}),
 where the coefficients that appear in the perturbation theory can be 
determined easily. 
One finds that the
attractor mechanism works to all orders without conditions other than
eq.(\ref{critical}), eq.(\ref{positive}) \footnote{For some specific values of the exponent $\gamma_i$,
eq.(\ref{order1}), though, we find that   there can be an obstruction  which prevents
 the solution from being extended to 
all orders.}.

When we turn
to other cases later in the paper, higher dimensional or AdS space
etc., we will sometimes not have explicit solutions, but an analysis
along the above lines in the near horizon and asymptotic regions and
showing regularity in-between will suffice to show that a smoothly
interpolating solution exists which connects the asymptotically flat
region to the attractor geometry at horizon.

To conclude, the key feature that leads to the attractor is the fact
that both solutions to the linearised equation for $\delta \phi$ are
well behaved as $r \rightarrow \infty$, and one solution near the
horizon is well behaved and vanishes.  If one of these features fails
the attractor mechanism typically does not work.  For example, adding
a mass term for the scalars results in one of the two solutions at
infinity diverging.  Now it is typically not possible to match the
well behaved solution near the horizon to the well behaved one at
infinity and this makes it impossible to turn on the dilaton
perturbation in a non-singular fashion.

We turn to a more detailed description of perturbation theory below.

\subsubsection{First Order Solution}

We start with first order perturbation theory.  We can write,
\begin{equation}
  \label{defepsilon}
  \delta \phi_i \equiv  \phi_i-\phi_{i0}=\epsilon\phi_{i1}, 
\end{equation}
where $\epsilon$ is the small parameter we use to organise the
perturbation theory.  The scalars $\phi_i$ are chosen to be
eigenvectors of the second derivative matrix, eq.(\ref{massmatrix}).

From, eq.(\ref{eq1}), eq.(\ref{eq2}), eq.(\ref{constraint}), we see
that there are no first order corrections to the metric components,
$a,b$. These receive a correction starting at second order in
$\epsilon$.  The first order correction to the scalars $\phi_i$
satisfies the equation,
\begin{eqnarray}
\label{eodf}
  \partial_{r}(a_0^{2}b_0^{2}\partial_{r}\phi_{i1})
  =\frac{\kkappa_i^{2}}{b_0^{2}}\phi_{i1}.
\end{eqnarray}
where, $\kkappa_i^2$ is the eigenvalue for the matrix
eq.(\ref{massmatrix}) corresponding to the mode $\phi_i$.
Substituting for $a_0,b_0,$ from eq.(\ref{ern}) we find,
\begin{equation}
  \phi_{i1}=c_{1i}\left(\frac{r-r_H}{r}\right)^{\frac{1}{2}(\pm \sqrt{1+4\kkappa_i^{2}/r_H^2}-1)}
\label{order1}
\end{equation}
We are interested in a solution which does not blow up at the horizon,
$r=r_H$. This gives,
\begin{equation}
  \phi_{i1}=c_{1i}\left(\frac{r-r_H}{r}\right)^{\gamma_i}\label{soldil1},
\end{equation}
where
\begin{eqnarray}
  \label{deft}
  \gamma_i&=&\textstyle\frac{1}{2}\left(\sqrt{1+\frac{4 \kkappa_i^{2}}{r_H^2}}-1\right).
\end{eqnarray}

Asymptotically, as $r \rightarrow \infty$, $\phi_{i1} \rightarrow
c_{1i}$, so the value of the scalars vary at infinity as $c_{1i}$ is
changed.  However, since $\gamma_{i}>0$, we see from
eq.(\ref{soldil1}) that $\phi_{i1}$ vanishes at the horizon and the
value of the dilaton is fixed at $\phi_{i0}$ regardless of its value
at infinity. This shows that the attractor mechanism works to first
order in perturbation theory.

It is worth commenting that the attractor behaviour arises because the   solution
to eq.(\ref{eodf}) which is non-singular at $r=r_H$, also vanishes there. 
To examine this further we write eq.(\ref{eodf})  in standard form,
\cite{MF}, 
\begin{equation}
\label{eq:standard:ode}
{d^2y \over dx^2} + P(x) y + Q(x) y=0,
\end{equation}
with $x=r-r_H$, $y=\phi_{i1}$.  The vanishing non-singular solution
arises because eq.(\ref{eodf}) has a single and double pole
respectively for $P(x)$ and $Q(x)$, as $x\rightarrow 0$. This results
in (\ref{eq:standard:ode}) having a scaling symmetry as $x\rightarrow
0$ and the solution goes like $x^{\gamma_i}$ near the horizon.  The
residues at these poles are such that the resulting indical equation
has one solution with exponent $\gamma_i>0$.  In contrast, in a
non-extremal black hole background, the horizon is still a regular
singular point for the  first order perturbation equation, but
$Q(x)$ has only a single pole.  It turns out that the resulting
non-singular solution can go to any constant value at the horizon and
does not vanish in general.

\subsubsection{Second Order Solution}

The first order perturbation of the dilaton sources a second order
correction in the metric.  We turn to calculating this correction
next.

Let us write,
\begin{eqnarray}
  b & = & b_{0}+\epsilon^{2}b_{2}\label{pert2}\\
  a^{2} & = & a_{0}^{2}+\epsilon^{2}a_{2}\nonumber \\
  b^{2} & = & b_{0}^{2}+2\epsilon^{2}b_{2}b_{0},\nonumber 
\end{eqnarray}
where $b_{0}$ and $a_{0}$ are the zeroth order extremal Reissner
Nordstrom solution eq.(\ref{ern}).

Equation (\ref{eq1}) gives,
\begin{equation}
  \label{sa1}
  a^2b^2=(r-r_H)^2 + d_1 r + d_2. 
\end{equation}
The two integration constants, $d_1,d_2$ can be determined by imposing
boundary conditions.  We are interested in extremal black hole
solutions with vanishing surface gravity. These should have a horizon
where $b$ is finite and $a^2$ has a ``double-zero'', i.e., both $a^2$
and its derivative $(a^2)'$ vanish. By a gauge choice we can always
take the horizon to be at $r=r_H$.  Both $d_1$ and $d_2$ then vanish.
Substituting eq.(\ref{pert2}) in the equation(\ref{eq1}) we get to
second order in $\epsilon$,
\begin{equation}
  2a_{0}^{2}b_{0}b_{2}+b_{0}^{2}a_{2}=0.
\end{equation}
Substituting for $a_0,b_0$ then determines, $a_2$ in terms of $b_2$,
\begin{equation}
  a_{2}=-2\left(1-\frac{r_{H}}{r}\right)^{2}\frac{b_{2}}{r}. \label{a2}
\end{equation}
From eq.(\ref{eq2}) we find next that,
\begin{equation}
  b_{2}(r)=-\sum_i \frac{c_{1i}^{2}\gamma}{2(2\gamma_i-1)}
  r \left(\frac{r-r_H}{r}\right)^{2\gamma_i} +A_{1}r +A_2r_H
  \label{b2}
\end{equation}
$A_1,A_2$ are two integration constants. The two terms proportional to
these integration constant solve the equations of motion for $b_2$ in
the absence of the $O(\epsilon)^2$ source terms from the dilaton.
This shows that the freedom associated with varying these constants is
a gauge degree of freedom.  We will set $A_1=A_2=0$ below.  Then,
$b_2$ is,
\begin{equation}
  b_{2}(r)=-\sum_i\frac{c_{1i}^{2}\gamma_i}{2(2\gamma_i-1)}
  r \left(\frac{r-r_H}{r}\right)^{2\gamma_i} 
  \label{finalb2}
\end{equation}
It is easy to check that this solves the constraint
eq.(\ref{constraint}) as well.

To summarise, the metric components to second order in $\epsilon$ are
given by eq.(\ref{pert2}) with $a_0,b_0$ being the extremal Reissner
Nordstrom solution and the second order corrections being given in
eq.(\ref{a2}) and eq.(\ref{finalb2}). Asymptotically, as $r
\rightarrow \infty$, $b_2 \rightarrow c \times r$, and,
$a_2\rightarrow -2 \times c$, so the solution continues to be
asymptotically flat to this order.  Since $\gamma_i>0$ we see from
eq.(\ref{a2}, \ref{finalb2}) that the second order corrections are
well defined at the horizon.  In fact since $b_2$ goes to zero at the
horizon, $a_2$ vanishes at the horizon even faster than a double-zero.
Thus the second order solution continues to be a double-zero horizon
black hole with vanishing surface gravity.  Since $b_2$ vanishes the
horizon area does not change to second order in perturbation theory
and is therefore independent of the asymptotic value of the dilaton.

The scalars also gets a correction to second order in $\epsilon$.
This can be calculated in a way similar to the above analysis.  We
will discuss this correction along with higher order corrections, in
one simple example, in the next subsection.

Before proceeding let us calculate the mass of the black hole to
second order in $\epsilon$.  It is convenient to define a new
coordinate,
\begin{equation}
  \label{by}
  y\equiv b(r)
\end{equation}
Expressing $a^2$ in terms of $y$ one can read off the mass from the
coefficient of the $1/y$ term as $y \rightarrow \infty$, as is
discussed in more detail in Appendix \ref{sec:appa}.  This gives,
\begin{equation}
  \label{mass2}
  M=r_H+\epsilon^{2}\sum_i\frac{r_Hc_{i1}^2\gamma_i}{2}
\end{equation}
where $r_H$ is the horizon radius given by (\ref{horizon}).  Since
$\gamma_i$ is positive, eq.(\ref{deft}), we see that as $\epsilon$
increases, with fixed charge, the mass of the black hole increases.
The minimum mass black hole is the extremal RN black hole solution,
eq.(\ref{ern}), obtained by setting the asymptotic values of the
scalars equal to their critical values.

\subsubsection{An Ansatz to All Orders}

Going to higher orders in perturbation theory is in principle
straightforward.  For concreteness we discuss the simple example,
eq.(\ref{se}), below.  We show in this example that the form of the
metric and dilaton can be obtained to all orders in perturbation
theory analytically.  We have not analysed the coefficients and
resulting convergence of the perturbation theory in great detail.  In
a subsequent section we will numerically analyse this example and find
that even the leading order in perturbation theory approximates the
exact answer quite well for a wide range of charges.  This discussion
can be generalised to other more complicated cases in a
straightforward way, although we will not do so here.

Let us begin by noting that eq.(\ref{eq1}) can be solved in general to
give,
\begin{equation}
  a^2b^2=(r-r_H)^2 + d_1r +d_2
  \label{gena}
\end{equation}
As in the discussion after eq(\ref{sa1}) we set $d_1=d_2=0,$ since we
are interested in extremal black holes.  This gives,
\begin{equation}
  a^2b^2=(r-r_H)^2, 
  \label{finala}
\end{equation}
where $r_H$ is the horizon radius given by eq.(\ref{horizon}).  This
can be used to determine $a$ in terms of $b$.

Next we expand $b$, $\phi$ and $a^2$ in a power series in $\epsilon$,
\begin{eqnarray}
  b & = & b_0+\sum_{n=1}^{\infty}\epsilon^{n}b_{n} \label{pertb}\\
  \phi & = & \phi_0 + \sum_{n=1}^\infty \epsilon^{n}\phi_{n} \label{pertphi} \\
  a^2 &    = & a_0^2 + \sum_{n=1}^\infty \epsilon^n a_n \label{perta}
\end{eqnarray}
where $b_0$, $a_0$ are given by eq.(\ref{ern}) and $\phi_0$ is given
by eq.(\ref{attractorexample}).

The ansatz which works to all orders is that the $n^{th}$ order terms
in the above two equations take the form,
\begin{equation}
  \phi_{n}(r)=c_{n}\left(\frac{r-r_{H}}{r}\right)^{n\gamma}\label{diln}
\end{equation}
\begin{equation}
  b_{n}(r)=d_{n}r\left(\frac{r-r_{H}}{r}\right)^{n\gamma}\label{bn}, \end{equation}
and, 
\begin{equation}
  a_n=e_n \left({r-r_H \over r}\right)^{n\gamma+2}, \label{an}
\end{equation}
where $\gamma$ is given by eqs.(\ref{deft}) and in this case takes the
value,
\begin{equation}
  \label{valgammase}
  \gamma={1\over 2}\left(\sqrt{1-2\alpha_1\alpha_2} -1\right). 
\end{equation} 
The discussion in the previous two subsections is in agreement with
this ansatz.  We found $b_1=0$, and from eq.(\ref{finalb2}) we see
that $b_2$ is of the form eq.(\ref{bn}). Also, we found $a_1=0$ and
from eq.(\ref{a2}) $a_2$ is of the form eq.(\ref{an}). And from
eq.(\ref{soldil1}) we see that $\phi_1$ is of form eq.(\ref{diln}).
We will now verify that this ansatz consistently solves the equations
of motion to all orders in $\epsilon$.  The important point is that
with the ansatz eq.(\ref{diln}, \ref{bn}) each term in the equations
of motion of order $\epsilon^n$ has a functional dependence $({(r-r_H)
  \over r})^{2\gamma n}$.  This allows the equations to be solved
consistently and the coefficients $c_n, d_n$ to be determined.

Let us illustrate this by calculating $c_2$.  From eq.(\ref{eq2}) and
eq.(\ref{finala}) we see that the equation of motion for $\phi$ can be
written in the form,
\begin{eqnarray}
  2b(r)^{2}\partial_{r}((r-r_{H})^{2}\partial_{r}\phi) & = & e^{\alpha_{i}\phi}Q_{i}^{2}\alpha_{i}\label{eqphi}
\end{eqnarray}
To $O(\epsilon^2)$ this gives,
\begin{equation}
  \label{int2}
  \left(\frac{r-r_H}{r}\right)^{2\gamma}
  \left(
    2c_{2}(e^{\alpha_{i}\phi_{0}}Q_{i}^{2}\alpha_{i}^2-4r_H^{2}\gamma(1+2\gamma))
    +e^{\alpha_{i}\phi_{0}}Q_{i}^{2}\alpha_{i}^3c_{1}^{2}
  \right)=0 
\end{equation}
Notice that the term $(\frac{r-r_H}{r})^{2\gamma}$ has factored out.
Solving eq.(\ref{int2}) for $c_2$ we now get,
\begin{eqnarray}
  c_{2}
  &=& \ha c_1^2 (\alpha_1+\alpha_2)\frac{(\gamma+1)}{(3\gamma+1)}
  \label{e2}
\end{eqnarray}

More generally, as discussed in Appendix \ref{sec:appa}, working to
the required order in $\epsilon$ we can recursively find, $c_n,d_n,
e_n$.

One more comment is worth making here. We see from eq.(\ref{finalb2})
that $b_2$ blows up when when $\gamma=1/2$. Similarly we can see from
eq.(\ref{dk_ans}) that $b_n$ blows up when $\gamma={1\over n}$ for
$b_n$.  So for the values, $\gamma={1\over n},$ where $n$ is an
integer, our perturbative solution does not work.

Let us summarise. We see in the simple example studied here that a
solution to all orders in perturbation theory can be found.  $b$,
$\phi$ and $a^2$ are given by eq.(\ref{bn}), eq.(\ref{diln}) and
eq.(\ref{an}) with coefficients that can be determined as discussed in
Appendix \ref{sec:appa}.  In the solution, $a^2$ vanishes at $r_H$ so
it is the horizon of the black hole.  Moreover $a^2$ has a double-zero
at $r_H$, so the solution is an extremal black hole with vanishing
surface gravity.  One can also see that $b_n$ goes linearly with $r$
as $r \rightarrow \infty$ so the solution is asymptotically flat to
all orders. It is also easy to see that the solution is non-singular
for $r\geqslant r_H$.  Finally, from eq.(\ref{diln}) we see that $\phi_n=0$,
for all $n>0$, so all corrections to the dilaton vanish at the
horizon.  Thus the attractor mechanism works to all orders in
perturbation theory.  Since all corrections to $b$ also vanish at the
horizon we see that the entropy is uncorrected in perturbation theory.
This is in agreement with the general argument given after
eq.(\ref{BH}). Note that no additional conditions had to be imposed,
beyond eq.(\ref{critical}, \ref{positive}), which were already
appeared in the lower order discussion, to ensure the attractor
behaviour \footnote{In our discussion of exact solutions in section 4
  we will be interested in the case, $\alpha_1=-\alpha_2$.  From
  eq.(\ref{e2}, \ref{dk_ans}) we see that the expressions for $c_{2}$
  and $d_{3}$ become,
  \begin{eqnarray}
    c_{2} & = & 0\\
    d_{3} & = & 0\end{eqnarray}
  It follows that in the perturbation series for $\phi$ and $b$ only the 
  $c_{2n+1}$(odd) terms and  $d_{2n}$(even) terms are non-vanishing respectively.}. 

\section{Numerical Results}\label{sec:num}

\begin{figure}
  \begin{center}
    \epsfig{file=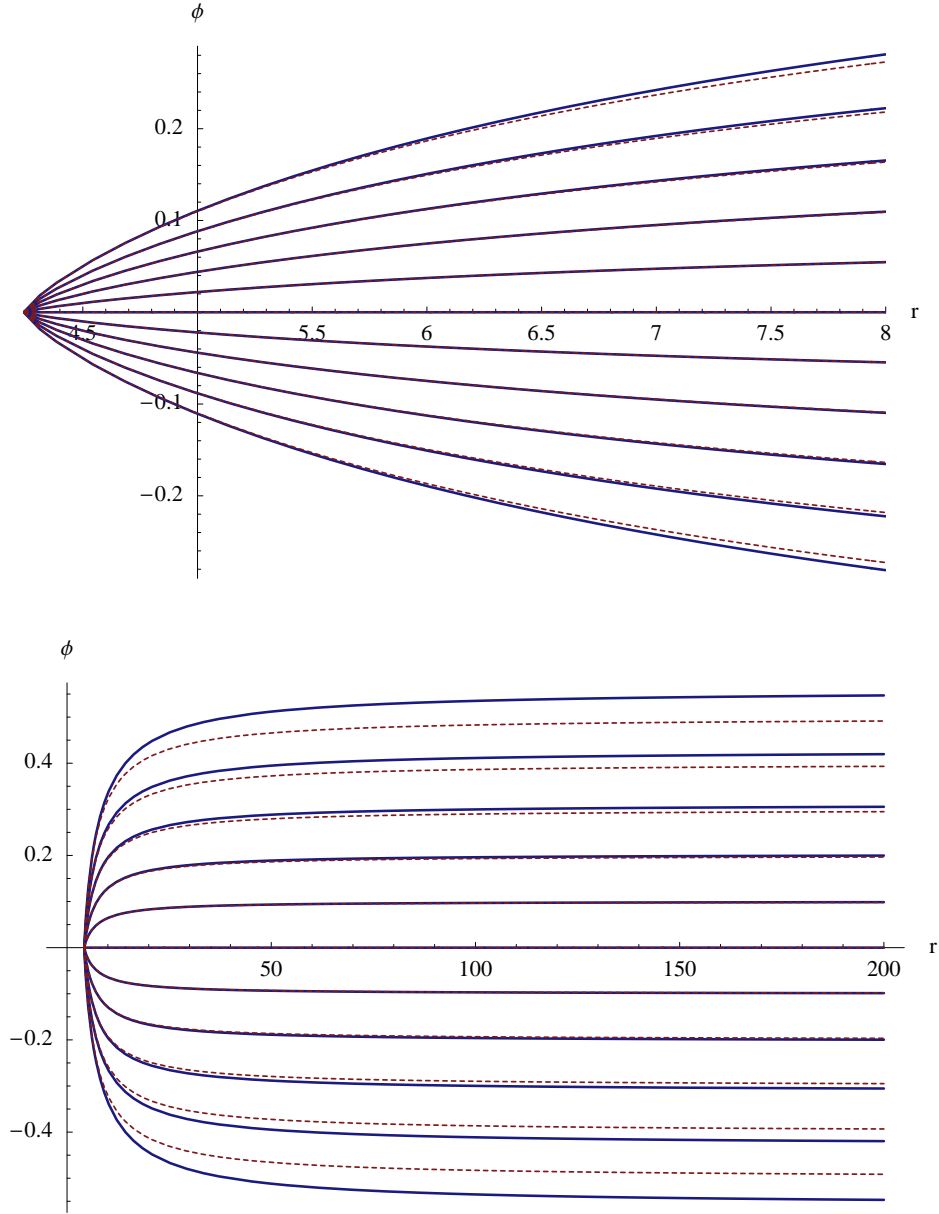,height=0.8\textheight}
  \end{center}
  \caption{\label{cap:num1}Comparison of numerical integration of
    $\phi$ with $1^{\mathrm{st}}$ order perturbation result. The upper
    graph is a close up of the lower one near the horizon. The
    perturbation result is denoted by a dashed line. We chose
    $\alpha_1,-\alpha_2=1.7$, $Q_1=3$, $Q_2=3$, $\delta r = 2.3\times
    10^{-8}$ and $c_1$ in the range $[-\frac{1}{2},\frac{1}{2}]$ }.
  $\phi_0 = 0 $.
\end{figure}
\begin{figure}
  \begin{center}
    \epsfig{file=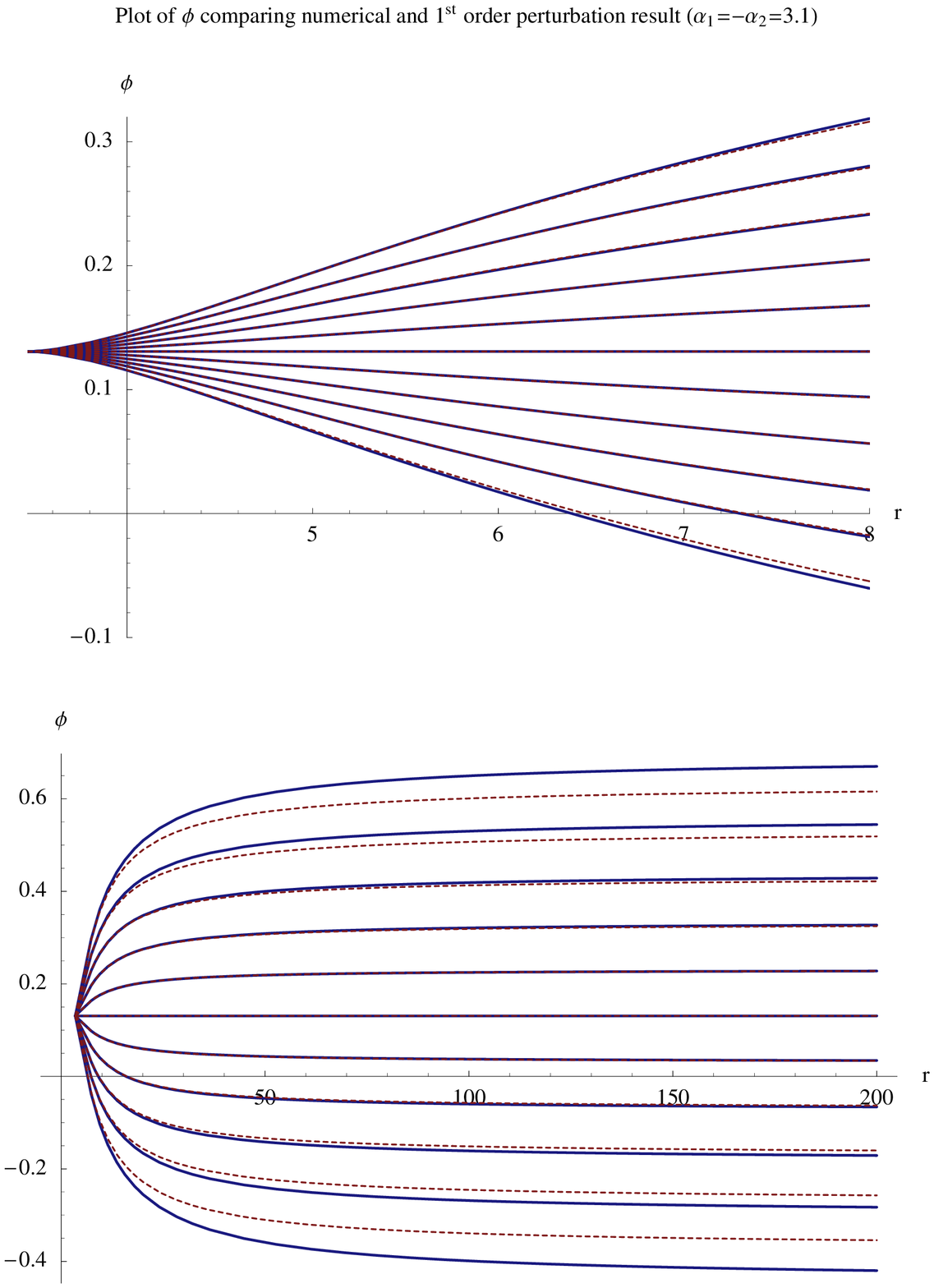,height=0.8\textheight}
  \end{center}
  \caption{\label{cap:num4}Comparison of numerical integration of
    $\phi$ with $1^{\mathrm{st}}$ order perturbation result. The upper
    graph is a close up of the lower one near the horizon. The
    perturbation result is denoted by a dashed line. We chose
    $\alpha_1,-\alpha_2=3.1$, $Q_1=2$, $Q_2=3$, $\delta r = 2.9\times
    10^{-8}$ and $c_1$ is in the range $[-\frac{1}{2},\frac{1}{2}]$ }.
  $\phi_0 = 0.13 $.
\end{figure}

There are two purposes behind the numerical work we describe in this
section.  First, to check how well perturbation theory works. Second,
to see if the attractor behaviour persists, even when $\epsilon$,
eq.(\ref{defepsilon}), is order unity or bigger so that the deviations
at asymptotic infinity from the attractor values are big.  We will
confine ourselves here to the simple example introduced near
eq.(\ref{se}), which was also discussed in the higher orders analysis
in the previous subsection.

In the numerical analysis it is important to impose the boundary
conditions carefully.  As was discussed above, the scalar has an
unstable mode near the horizon.  Generic boundary conditions imposed
at $r\rightarrow \infty$ will therefore not be numerically stable and
will lead to a divergence.  To avoid this problem we start the
numerical integration from a point $r_i$ near the horizon.  We see
from eq.(\ref{diln}, \ref{bn}) that sufficiently close to the horizon
the leading order perturbative corrections \footnote{We take the
  $O(\epsilon)$ correction in the dilaton, eq.(\ref{soldil1}), and the
  $O(\epsilon^2)$ correction in $b, a^2$, eq.(\ref{a2}, \ref{b2}).
  This consistently meets the constraint eq.(\ref{constraint}) to
  $O(\epsilon^2)$.}  becomes a good approximation.  We use these
leading order corrections to impose the boundary conditions near the
horizon and then numerically integrate the exact equations,
eq.(\ref{eq1},\ref{eq2}), to obtain the solution for larger values of
the radial coordinate.

The numerical integration is done using the Runge-Kutta method.  We
characterise the nearness to the horizon by the parameter
\begin{equation}
  \label{eq:def:deltar}
  \delta r = \frac{r_i-r_H}{r_i}
\end{equation}
where $r_i$ is the point at which we start the integration.  $c_1$
refers to the asymptotic value for the scalar, eq.(\ref{soldil1}).

In figs. (\ref{cap:num1},\ref{cap:num4}) we compare the numerical and
$1^\mathrm{st}$ order correction. The numerical and perturbation
results are denoted by solid and dashed lines respectively.  We see
good agreement even for large $r$.  As expected, as we increase the
asymptotic value of $\phi$, which was the small parameter in our
perturbation series, the agreement decreases.

Note also that the resulting solutions turn out to be singularity free
and asymptotically flat for a wide range of initial conditions.  In
this simple example there is only one critical point,
eq.(\ref{attractorexample}).  This however does not guarantee that the
attractor mechanism works. It could have been for example that as the
asymptotic value of the scalar becomes significantly different from
the attractor value no double-zero horizon black hole is allowed and
instead one obtains a singularity.  We have found no evidence for
this. Instead, at least for the range of asymptotic values for the
scalars we scanned in the numerical work, we find that the attractor
mechanism works with attractor value, eq.(\ref{attractorexample}).

It will be interesting to analyse this more completely, extending this
work to cases where the effective potential is more complicated and
several critical points are allowed. This should lead to multiple
basins of attraction as has already been discussed in the
supersymmetric context in e.g., \cite{Denefa, Denef:2001xn}.

\section{Exact Solutions }\label{sec:exact_sol}
\newcommand{\deltaa}{\alpha_1-\alpha_2}
\newcommand{\bovera}{\frac{\alpha_2}{\alpha_1}}
\newcommand{\fooa}{\frac{\alpha_1}{\deltaa}}
\newcommand{\foob}{\frac{-\alpha_2}{\deltaa}}

In certain cases the equation of motion can be solved exactly
\cite{Gibbons:1987ps}.  In this section, we shall look at some
solvable cases and confirm that the extremal solutions display
attractor behaviour. In particular, we shall work in $4$ dimensions
with one scalar and two gauge fields, taking $V_{eff}$ to be given by
eq.(\ref{seffpot}),
\begin{equation}
  \label{seffpot2}
  V_{eff}=e^{\alpha_1 \phi} (Q_1)^2 + e^{\alpha_2 \phi} (Q_2)^2.
\end{equation} 
We find that at the horizon the scalar field relaxes to the attractor
value (\ref{attractorexample})
\begin{equation}
  e^{(\alpha_1-\alpha_2)\phi_{0}}
  =-\frac{\alpha_2 Q_{2}^{2}}{\alpha_1 Q_{1}^{2}}\label{eq:attractor_again}
\end{equation}
which is the critical point of $V_{eff}$ and independent of the
asymptotic value, $\phi_\infty$.  Furthermore, the horizon area is
also independent of $\phi_\infty $ and, as predicted in section
\ref{sec:cond:attr}, it is proportional to the effective potential
evaluated at the attractor point.  It is given by
\begin{eqnarray}
  \mathrm{Area} &=& 4\pi b_H^2 = 4\pi V_{eff}(\phi_0) \\
  &=& 4\pi \spadesuit (Q_1)^{2\foob}  (Q_2)^{2\fooa}\label{eq:horizon_area}
\end{eqnarray} 
where
\begin{equation}
  \spadesuit = \textstyle 
  \left(-\bovera\right)^\fooa +
  \left(-\bovera\right)^\foob
  \label{eq:def_spade}
\end{equation}
is a numerical factor. It is worth noting that when
$\alpha_1=-\alpha_2$, one just has
\begin{equation}
  \label{eq:simple_area}
  \textstyle{1\over4}\mathrm{Area}= 2\pi  |Q_1 Q_2|
\end{equation}

Interestingly, the solvable cases we know correspond to $\gamma=1,2,3$
where $\gamma$ is given by (\ref{deft}). The known solutions for
$\gamma=1,2$ are discussed in \cite{Gibbons:1987ps} and references
therein (although they fixed $\phi_\infty=0$). We found a solution for
$\gamma=3$ and it appears as though one can find exact solutions as
long as $\gamma$ is a positive integer. Details of how these solutions
are obtained can be found in the references and appendix
\ref{sec:details}. 
 
For the cases we consider, the extremal solutions can be written in
the following form
\begin{eqnarray}
  e^{(\alpha_1-\alpha_2)\phi} & = & \left(-\frac{\alpha_2}{\alpha_1}\right)
  \left(\frac{\PP}{\Q}\right)^2
  \left(\frac{f_2}{f_1}\right)^{-\ha\alpha_1\alpha_2}\label{eq:exact_scalar_sol}\\
  b^{2} & = &
  \spadesuit \left((Q_1 f_1)^{-\alpha_2}  (Q_2 f_2)^{\alpha_1}\right)^{{\frac{2}{\deltaa}}}
  \label{eq:exact_b_sol}\\
  a^{2} &=& \rho^2/b^2
\end{eqnarray}
where $\rho=r-r_H$ and the $f_i$ are polynomials in $\rho$ to some
fractional power. In general the $f_i$ depend on $\phi_\infty$ but
they have the property
\begin{equation}
  \label{eq:poly_prop}
  f_i|_\mathrm{Horizon}=1.
\end{equation}
Substituting (\ref{eq:poly_prop}) into
(\ref{eq:exact_scalar_sol},\ref{eq:exact_b_sol}), one sees that that
at the horizon the scalar field takes on the attractor value
(\ref{eq:attractor_again}) and the horizon area is given by
(\ref{eq:horizon_area}).

\begin{figure}
  \begin{center}
    \epsfig{file=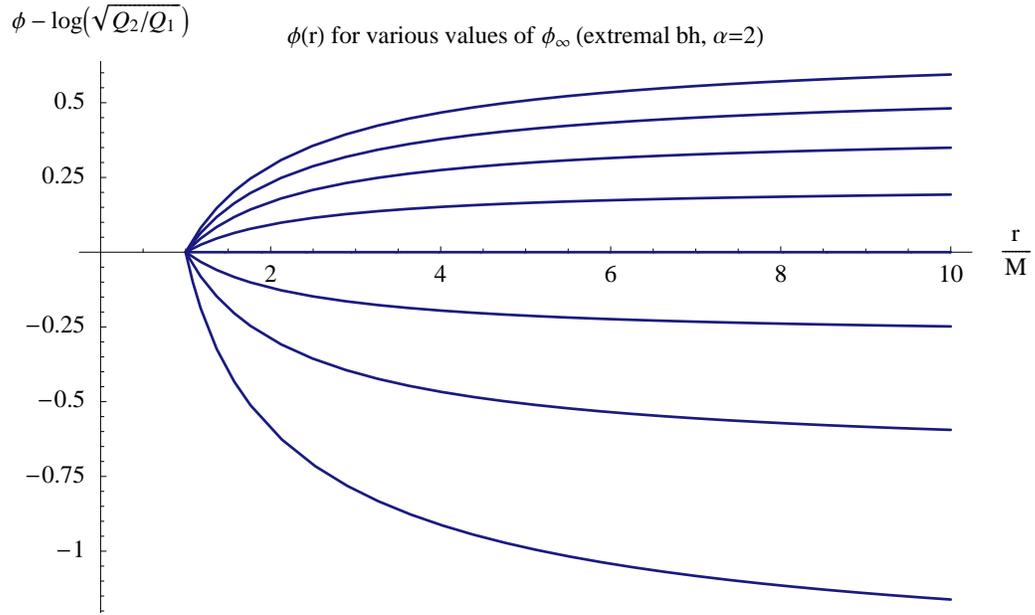,width=\linewidth}
  \end{center}
  \caption{\label{cap:nice_graphs}Attractor behaviour for the case
    $\gamma=1$; $\alpha_1,-\alpha_2=2$}
\end{figure}
\begin{figure}
  \begin{center}
    \epsfig{file=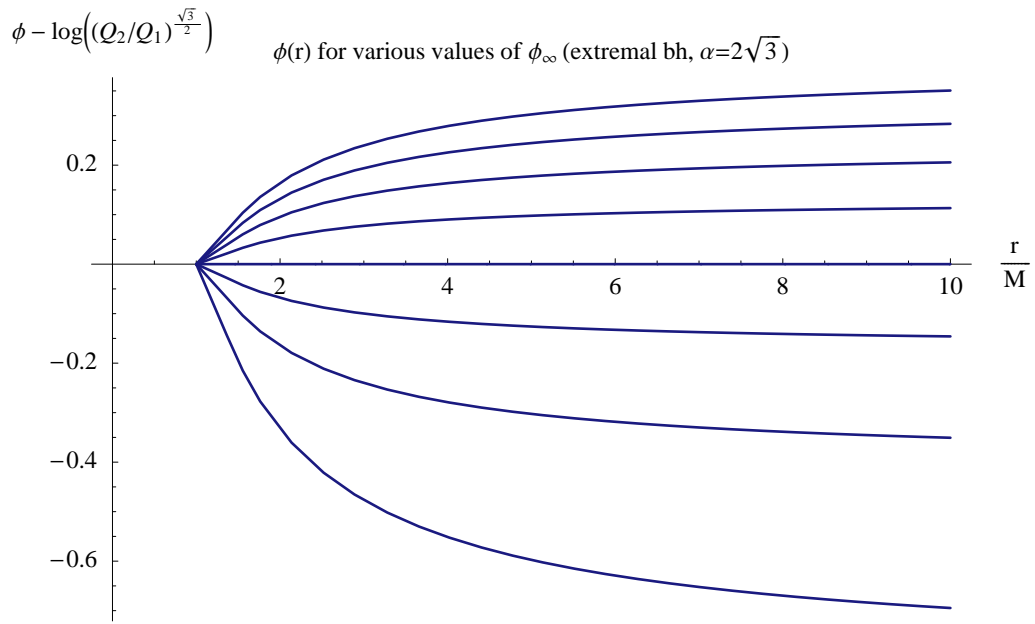,width=\linewidth}
  \end{center}
  \caption{\label{cap:nice_graphs2}Attractor behaviour for the case
    $\gamma=2$; $\alpha_1,-\alpha_2=2\sqrt{3}$}
\end{figure}

Notice that, when $\alpha=|\alpha_i|$,
(\ref{eq:exact_scalar_sol},\ref{eq:exact_b_sol}) simplify to
\begin{eqnarray}
  e^{\alpha\phi} & = & \frac{|\PP|}{|\Q|}\left(\frac{f_2} {f_1}\right)^{\frac{1}{4}\alpha^2}\\
  b^{2} & = & 2 |Q_1| |Q_2| \left( f_1  f_2\right)
\end{eqnarray}

\subsection{Explicit Form of the $f_i$}
\label{sec:explicit}

In this section we present the form of the functions $f_i$ mainly to
show that, although they depend on $\phi_\infty$ in a non trivial way,
they all satisfy (\ref{eq:poly_prop}) which ensures that the attractor
mechanism works. It is convenient to define
\begin{equation}
  \bar{Q}_i^2 = e^{\alpha_i\phi_\infty}Q_i^2\nosum
  \label{eq:def_barQ}
\end{equation}
which are the effective $U(1)$ charges as seen by an asymptotic
observer. For the simplest case, $\gamma=1$, we have
\begin{equation}
  f_{i}=1+\textstyle\left(\bar{Q}_i^{-1}|\alpha_i|{(4+\alpha_i^2)}^{-\ha}\right)\rho
\end{equation}
Taking $\gamma=2$ and $\alpha_1=-\alpha_2=2\sqrt{3}$ one finds
\begin{equation}
  f_i  = 
  \left( 
    1+
    (\bar{Q}_1 \bar{Q}_2)^{-\frac{2}{3}} (\bar{Q}_1^{\frac{2}{3}} + \bar{Q_2}^{\frac{2}{3}})^\ha \rho
    +{\textstyle\ha} (\bar{Q}_i\bar{Q}_1 \bar{Q}_2)^{-\frac{2}{3}}\rho^{2}
  \right)^\ha
\end{equation}
Finally for $\gamma=3$ and $\alpha_1=4$ $, \alpha_2=-6$ we have
\begin{equation}
  f_{1} =  
  \left(
    1-6a_{2}\rho + 12a_{2}^{2}\rho^{2}-6a_{0}\rho^{3}
  \right)^{\frac{1}{3}}
\end{equation}
\begin{equation}
  f_{2}  =  
  \left(
    1-
    {\textstyle \frac{24}{3}}a_{2}\rho
    +24a_{2}\rho^{2}
    -(48a_{2}^{3}-12a_{0})\rho^{3}
    +\left(48a_{2}^{4}-24a_{0}a_{2}\right)\rho^{4}
  \right)^{\frac{1}{4}}
\end{equation}
where $a_0$ and $a_2$ are non-trivial functions of $\bar{Q}_i$.  Further details are
discussed in section \ref{sec:nonex} and appendix \ref{sec:details}. The scalar field
solutions for $\gamma=1$ and $2$ are illustrated in figs. \ref{cap:nice_graphs} and
\ref{cap:nice_graphs2} respectively.

\subsection{Supersymmetry and the Exact Solutions}

As mentioned above, the first two cases ($\gamma=1,2$) have been
extensively studied in the literature.

The SUSY of the extremal $\alpha_1=-\alpha_2=2$ solution is discussed
in \cite{Kallosh:1992ii}. They show that it is supersymmetric in the
context of ${\cal N}=4$ SUGRA. It saturates the BPS bound and
preserves $1\over4$ of the supersymmetry - ie. it has ${\cal N}=1$
SUSY. There are BPS black-holes in this context which carry only one
$U(1)$ charge and preserve $\ha$ of the supersymmetry. The
non-extremal blackholes are of course non-BPS.

On the other hand, the extremal $\alpha_1=-\alpha_2=2\sqrt3$ blackhole
is non-BPS \cite{Gibbons:1994ff}. It arises in the context of
dimensionally reduced $5D$ Kaluza-Klein gravity \cite{Dobiasch:1981vh}
and is embeddable in ${\cal N}=2$ SUGRA. There however are BPS
black-holes in this context which carry only one $U(1)$ charge and
once again preserve $\ha$ of the supersymmetry \cite{Gibbons:1993xt}.

We have not investigated the supersymmetry of the $\gamma=3$ solution,
we expect that it is not a BPS solution in a supersymmetric theory.

\section{General Higher Dimensional Analysis}
\subsection{The Set-Up}

It is straightforward to generalise our results above to higher
dimensions.  We start with an action of the form,
\begin{equation}
  \label{higherdaction}
  S=\frac{1}{\kappa^{2}}\int d^{d}x\sqrt{-G}(R-2(\partial\phi_i)^{2}-
  f_{ab}(\phi_i)F^a F^b)
\end{equation}
Here the field strengths, $F_a$ are $(d-2)$ forms which are magnetic
dual to $2$-form fields.

We will be interested in solution which preserve a $SO(d-2)$ rotation
symmetry. Assuming all quantities to be function of $r$, and taking
the charges to be purely magnetic, the ansatz for the metric and gauge
fields is \footnote{Black hole which carry both electric and magnetic
  charges do not have an $SO(d-2)$ symmetry for general $d$ and we
  only consider the magnetically charged case here. The analogue of
  the two-form in 4 dimensions is the $d/2$ form in $d$ dimensions. In
  this case one can turn on both electric and magnetic charges
  consistent with $SO(d/2)$ symmetry. We leave a discussion of this
  case and the more general case of $p$-forms in $d$ dimensions for
  the future.}
\begin{eqnarray}
  \label{metrichd}
  ds^{2}=-a(r)^{2}dt^{2}+a(r)^{-2}dr^{2}+b(r)^{2}d\Omega_{d-2}^{2}\\
  F^a=Q^a\sin^{d-3}\theta\sin^{d-4}\phi\cdots d\theta\wedge d\phi\wedge\cdots\\
  \tilde{F}^a=Q^{a}\sin^{d-3}\theta\sin^{d-4}\phi\cdots d\theta\wedge d\phi\wedge\cdots
\end{eqnarray}

The equation of motion for the scalars is
\begin{equation}
  \partial_{r}(a^{2}b^{d-2}\partial_{r}\phi_i)  = \frac{(d-2)!
    \partial_iV_{eff}} {4b^{d-2}}.
  \label{dilatoneqhd}
\end{equation}
Here $V_{eff}$, the effective potential for the scalars, is given by
\begin{equation}
  \label{eact}
  V_{eff}=f_{ab}(\phi_i) Q^aQ^b.
\end{equation}

From the $(R_{rr}-{G^{tt} \over G^{rr}} R_{tt})$ component of the
Einstein equation we get,
\begin{eqnarray}
  \sum_i(\phi_i')^{2}= -{(d-2) b''(r) \over 2 b(r)}.
  \label{Rrreq}
\end{eqnarray}
The $R_{rr}$ component gives the constraint,
\begin{equation}
  \label{constrainthd}
  \begin{array}{l}
    -(d-2)\{ (d-3)-ab'(2a'b+(d-3)ab')\}     
    =  2\phi_i'^{2}a^{2}b^{2} -\frac{(d-2)!}{b^{2(d-3)}} V_{eff}(\phi_i)
  \end{array}
\end{equation}

In the analysis below we will use eq.(\ref{dilatoneqhd}) to solve for
the scalars and then eq.(\ref{Rrreq}) to solve for $b$. The constraint
eq.(\ref{constrainthd}) will be used in solving for $a$ along with one
extra relation, $R_{tt}=(d-3) {a^2 \over b^2} R_{\theta\theta}$, as is
explained in appendix \ref{sec:higherdim}.  These equations (aside
from the constraint) can be derived from a one-dimensional action
\begin{equation}
  \label{aohdv}
  \begin{array}{cccc}
    S&=&\frac{1}{\kappa^{2}}\int dr &
    \Big((d-3)(d-2)b^{d-4}(1+a^{2}b^{'2})+(d-2)b^{d-3}(a^{2})^{'}b^{'} \\
    &&&   -2a^{2}b^{d-2}(\partial_{r
    }\phi)^{2}-\frac{(d-2)!}{b^{d-2}}V_{eff}\Big)
  \end{array}
\end{equation}

As the analysis below shows if the potential has a critical point at
$\phi_i=\phi_{i0}$ and all the eigenvalues of the second derivative
matrix $\partial_{ij}V(\phi_{i0})$ are positive then the attractor
mechanism works in higher dimensions as well.

\subsection{Zeroth and First Order Analysis}
Our starting point is the case where the scalars take asymptotic
values equal to their critical value, $\phi_i=\phi_{i0}$. In this case
it is consistent to set the scalars to be a constant, independent of
$r$.  The extremal Reissner Nordstrom black hole in $d$ dimensions is
then a solution of the resulting equations.  This takes the form,
\begin{eqnarray}
  \label{ernhd}
  a_0(r)=\left(1-\frac{r_{H}^{d-3}}{r^{d-3}}\right)\quad
  b_0(r)=r
\end{eqnarray}
where $r_{H}$ is the horizon radius.  From the eq.(\ref{constrainthd})
evaluated at $r_H$ we obtain the relation,
\begin{eqnarray}
  r_{H}^{2(d-3)} =  (d-4)! V_{eff}(\phi_{i0})
\end{eqnarray}
Thus the area of the horizon and the entropy of the black hole are
determined by the value of $V_{eff}(\phi_{i0})$, as in the
four-dimensional case.

Now, let us set up the first order perturbation in the scalar fields,
\begin{equation}
  \phi_i=\phi_{i0}+\epsilon\phi_{i1}
\end{equation}
The first order correction satisfies,
\begin{eqnarray}
  \partial_{r}(a_0^{2}b_0^{d-2}\partial_{r}\phi_{i1})=\frac{\kkappa_i^{2}}{b^{d-2}}\phi_{i1}
\end{eqnarray}
where, $\kkappa_i^2$ is the eigenvalue of the second derivative matrix
${(d-2)! \over 4} \partial_{ij}V_{eff}(\phi_{i0})$ corresponding to
the mode $\phi_i$.  This equation has two solutions. If $\kkappa_i^2
>0$ one of these solutions blows up while the other is well defined
and goes to zero at the horizon.  This second solution is the one we
will be interested in. It is given by,
\begin{equation}
  \phi_{i1}=c_{i1}(1-{r_{H}^{d-3}/r^{d-3}})^{\gamma_i}
\end{equation}
where $\gamma$ is given by
\begin{eqnarray}
  \gamma_i={\frac{1}{2}}\left(-1+{\sqrt{1+4\kkappa_i^{2}r_{H}^{6-2d}/(d-3)^{2}}}\right)
\end{eqnarray}

\subsubsection{Second order calculations (Effects of backreaction)}

The first order perturbation in the scalars gives rise to a second
order correction for the metric components, $a,b$.  We write,
\begin{eqnarray}
  b(r)=b_{0}(r)+\epsilon^{2}b_{2}(r)\label{pertstart}\\
  a(r)^{2}=a_{0}(r)^{2}+\epsilon^{2}a_{2}(r)\\
  b(r)^{2}=b_{0}(r)^{2}+2\epsilon^{2}b_{2}(r)b_{0}(r)\label{pertend}
\end{eqnarray}
where $a_0,b_{0}$ are given in eq.(\ref{ernhd}).

From (\ref{Rrreq}) one can solve for the second order perturbation
$b_{2}(r)$. For simplicity we consider the case of a single scalar
field, $\phi$. The solution is given by double-integration form,
\begin{eqnarray}
  {\partial_{r}^{2}b_{2}(r)} & = & 
  -{\frac{2}{(d-2)}}r(\partial_{r}\phi_{1})^{2}
  =-c_{1}'{\frac{1}{r^{2d-5}}}({\frac{{r^{d-3}-r_{H}^{d-3}}}{{r^{d-3}}}})^{2\gamma-2}\nonumber \\
  \Rightarrow b_{2}(r) & = &  d_{1}r+d_{2}
  -{\frac{c_{1}' r}{{2(d-3)(d-4)\gamma(2 \gamma-1)r_H^{2d}}}}\times \nonumber\\
  && \Bigl(-(d-4)
  F\left[{\textstyle{\frac{1}{{3-d}}},1-2\gamma,{\frac{{d-4}}{{d-3}}};({\frac{r_H}{r}})^{d-3}}\right]\nonumber \\
  & & +(2\gamma-1){\textstyle \left(\frac{r_H}{r}\right)^{d-3}}F\left[{\textstyle{\frac{{d-4}}{{d-3}}},1-2\gamma,{\frac{{2d-7}}{{d-3}}};({\frac{r_H}{r}})^{d-3}}\right]\Bigr)\,,
  \label{btwoexact}
\end{eqnarray}
where $c_{1}'\equiv2(d-3)^{2}c_{1}^{2}\gamma^{2}r_H^{d}/(d-2)$, a
positive definite constant, and $F$ is Gauss's Hypergeometric
function.  More generally, for several scalar fields, $b_2$ is
obtained by summing over the contributions from each scalar field.
The integration constants $d_1, d_2$, in eq.(\ref{btwoexact}), can be
fixed by coordinate transformations and requiring a double-zero
horizon solution. We will choose coordinate so that the horizon is at
$r=r_H$, then as we will see shortly the extremality condition
requires both $d_1,d_2$ to vanish.  As $r \rightarrow r_H$ we have
from eq.(\ref{btwoexact}) that
\begin{equation}
  b_{2}(r)\propto-\left({\frac{{r^{d-3}-r_{H}^{d-3}}}{{r^{d-3}}}}\right)^{2\gamma}
\end{equation}
Since $\gamma>0$, we see that $b_2$ vanishes at the horizon and thus
the area and the entropy are uncorrected to second order.  At large
$r$,
$b_{2}(r)\propto{\mathcal{O}}(r)+{\mathcal{O}}(1)+{\mathcal{O}}(r^{7-2d})$
so asymptotic behaviour is consistent with asymptotic flatness of the
solution.

The analysis for $a_2$ is discussed in more detail in appendix
\ref{sec:higherdim}.  In the vicinity of the horizon one finds that
there is one non-singular solution which goes like, $a_2(r)
\rightarrow C (r-r_H)^{(2\gamma +2)}$.  This solution smoothly extends
to $r \rightarrow \infty$ and asymptotically, as $r\rightarrow
\infty$, goes to a constant which is consistent with asymptotic
flatness.

Thus we see that the backreaction of the metric is finite and well
behaved.  A double-zero horizon black hole continues to exist to
second order in perturbation theory. It is asymptotically flat.  The
scalars in this solution at the horizon take their attractor values
irrespective of their values at infinity..

Finally, the analysis in principle can be extended to higher orders.
Unlike four dimensions though an explicit solution for the higher
order perturbations is not possible and we will not present such an
higher order analysis here.

We end with Fig 5.  which illustrates the attractor behaviour in
asymptotically flat $4+1$ dimensional space. This figure has been
obtained for the example, eq.(\ref{se}, \ref{seffpot}). The parameter
$\delta r$ is defined in eq.(\ref{eq:def:deltar}).
 
\begin{figure}[htb]
  \begin{center}
    \epsfig{file=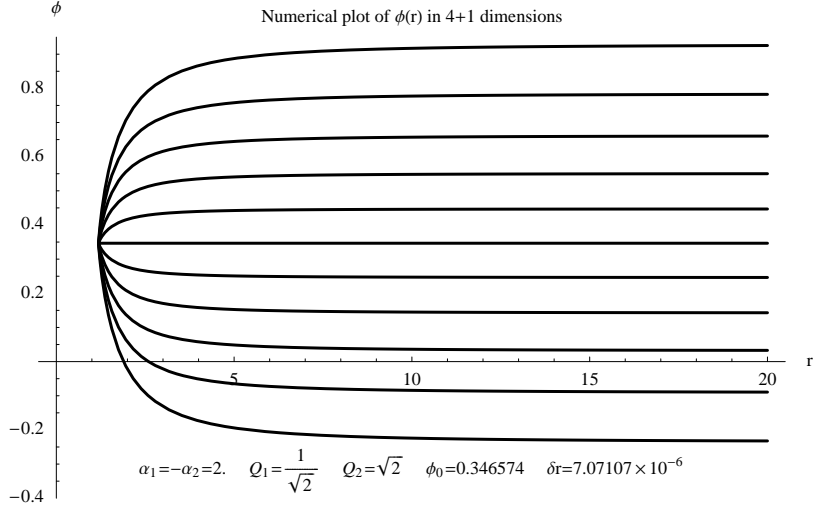,width=0.8\linewidth}
  \end{center}
  \caption{\label{cap:$4+1$ Dimensional Attractor}Numerical plot of
    $\phi(r)$ with $\alpha_1=-\alpha_2=2$ for the an extremal black
    hole in $4+1$ dimensions displaying attractor behaviour.}
\end{figure}

\section{Attractor in $AdS_4$}
Next we turn to the case of Anti-de Sitter space in four dimensions.
Our analysis will be completely analogous to the discussion above for
the four and higher dimensional case and so we can afford to be
somewhat brief below.

The action in 4-dim. has the form
\begin{equation}
  \label{fourdact}
  S=\frac{1}{\kappa^{2}}\int d^{4}x\sqrt{-G}
  (R-2\Lambda-2(\partial\phi_i)^{2}-f_{ab}(\phi_i) F^a F^b
  -{\textstyle {1 \over 2}} {\tilde f}_{ab}
  (\phi_i) \epsilon^{\mu \nu \rho \sigma} F^a_{\mu \nu} F^b_{\rho \sigma} ) 
\end{equation}
where $\Lambda=-3/L^{2}$ is the cosmological constant.  For simplicity
we will discuss the case with only one scalar field here. The
generalisation to many scalars is immediate and along the lines of the
discussion for asymptotically flat four-dimensional case. Also we take
the coefficient of the scalar kinetic energy term to be field
independent.

For spherically symmetric solutions the metric takes the form,
eq.(\ref{metric2}). The field strengths are given by
eq.(\ref{fstrenghtgen}).  This gives rise to a one dimensional action
\begin{equation}
  \label{actoned}
  S=\frac{1}{\kappa^{2}}\int dr
  \left(2-(a^{2}b^{2})^{''}-2a^{2}bb^{''}-2a^{2}b^{2}(\partial_{r}\phi)^{2}-2\frac{V_{eff}}{b^{2}}
    +\frac{3b^{2}}{L^{2}}\right),
\end{equation} 
where $V_{eff}$ is given by eq.(\ref{defpotgen}).  The equations of
motion, which can be derived either from eq.(\ref{actoned}) or
directly from the action, eq.(\ref{fourdact}) are now given by,
\begin{eqnarray}
  \partial_{r}(a^{2}b^{2}\partial_{r}\phi)=\frac{\partial_\phi V_{eff}(\phi)}{2b^{2}}\\
  \frac{b^{''}}{b}=-(\partial_{r}\phi)^{2},\label{eq2ads}\end{eqnarray}
which are unchanged from the flat four-dimensional case,
and,
\begin{equation}
  (a^{2}(r)b^{2}(r))^{''}=2(1-2\Lambda b^{2}),\label{eq1ads}\end{equation}
\begin{equation}
  -1+a^{2}b^{'2}+\frac{a^{2'}b^{2'}}{2}=\frac{-1}{b^{2}}(V_{eff}(\phi))+a^{2}b^{2}(\partial_{r}\phi)^{2}+\frac{3b^{2}}{L^2},\label{constraintads}\end{equation}
where the last equation is the first order constraint.

\subsection{Zeroth and First Order Analysis for $V$}
The zeroth order solution is obtained by taking the asymptotic values
of the scalar field to be its critical values, $\phi_{0}$ such that
$\partial_iV_{eff}(\phi_{0})=0$.

The resulting metric is now the extremal Reissner Nordstrom black hole
in AdS space, \cite{Chamblin:1999tk}, given by,
\begin{eqnarray}
  a_{0}(r)^{2}&=&{\frac{{(r-r_{H})^{2}(L^2+3r_{H}^{2}+2r_{H}r+r^{2})}}{{L^2r^{2}}}}\\
  b_{0}(r)&=&r\end{eqnarray}

The horizon radius $r_H$ is given by evaluating the constraint
eq.(\ref{constraintads}) at the horizon,
\begin{eqnarray}
  {\frac{{(L^2r_H^{2}+2r_H^{4})}}{{L^2}}}=V_{eff}(\phi_{0}).\nonumber \\
\end{eqnarray}

The first order perturbation for the scalar satisfies the equation,
\begin{eqnarray}
  \label{seads}
  \partial_{r}(a_{0}^{2}b_0^{2}\partial_{r}\phi_{1})=\frac{\kkappa^{2}}{b^{2}}\phi_{1}\end{eqnarray}
where,
\begin{equation}
  \label{valkappaads}
  \kkappa^2={1 \over 2}\partial_\phi^2V_{eff}(\phi_0).
\end{equation}
This is difficult to solve explicitly.
 
In the vicinity of the horizon the two solutions are given by
\begin{equation}
  \label{sfoads}
  \phi_1=C_{\pm} (r-r_H)^{t_{\pm}}
\end{equation}
If $V_{eff}''(\phi_0)>0$ one of the two solutions vanishes at the
horizon.  We are interested in this solution. It corresponds to the
choice,
\begin{equation}
  \label{sfoadsb}
  \phi_1=C (r-r_H)^\gamma,
\end{equation}
where,
\begin{equation}
  \label{valtpads}
  \gamma=
  \frac{\sqrt{1+\frac{4\kkappa^{2}}{\delta r_H^{2}}}-1}{2},
\end{equation}
and, $\delta={(L^2+6r_H^2)\over L^2}$.  As discussed in the appendix
\ref{sec:ads} this solution behaves at $r \rightarrow \infty$ as
$\phi_1 \rightarrow C_1 +C_2/r^3$.  Also, all other values of $r$,
besides the horizon and $\infty$, are ordinary points of the second
order equation eq.(\ref{seads}). All this establishes that there is
one well-behaved solution for the first order scalar perturbation. In
the vicinity of the horizon it takes the form eq.(\ref{sfoads}) with
eq.(\ref{valtpads}), and vanishes at the horizon. It is non-singular
everywhere between the horizon and infinity and it goes to a constant
asymptotically at $r \rightarrow \infty$.

We consider metric corrections next.  These arise at second order. We
define the second order perturbations as in eq.(\ref{pert2}).  The
equation for $b_2$ from the second order terms in eq.(\ref{eq2ads})
takes the form,
\begin{eqnarray}
  b_{2}''=-r(\phi_{1}'(r))^{2},\label{adsb2eq}
\end{eqnarray}
and can be solved to give,
\begin{equation}
  \label{b2ads}
  b_2(r)=-\int_{r_H}\int_{r_H} [r (\phi_{1}'(r))^{2}] 
\end{equation}
We fix the integration constants by taking take the lower limit of
both integrals to be the horizon. We will see that this choice gives
rise to an double-zero horizon solution. Since $\phi_{1}$ is well
behaved for all $r_H\le r \le \infty$ the integrand above is well
behaved as well. Using eq.(\ref{sfoads}) we find that in the near
horizon region
\begin{equation}
  \label{nhb2ads}
  b_2 \sim (r-r_H)^{(2 \gamma)}
\end{equation}

At $r \rightarrow \infty$ using the fact that $\phi_1 \rightarrow C_1
+C_2/r^3$ we find
\begin{equation}
  \label{asbads}
  b_2 \sim D_1 r + D_2 + D_3/r^6.
\end{equation}
This is consistent with an asymptotically AdS solution.

Finally we turn to $a_2$. As we show in appendix \ref{sec:ads} a
solution can be found for $a_2$ with the following properties.  In the
vicinity of the horizon it goes like,
\begin{equation}
  \label{nhads}
  a_2 \propto (r-r_H)^{(2\gamma + 2)},
\end{equation}
and vanishes faster than a double-zero.  As $r \rightarrow \infty$,
$a_2 \rightarrow d_1 r $ and grows more slowly than $a_0^2$.  And for
$r_H<r<\infty$ it is well-behaved and non-singular.

This establishes that after including the backreaction of the metric
we have a non-singular, double-zero horizon black hole which is
asymptotically AdS. The scalar takes a fixed value at the horizon of
the black hole and the entropy of the black hole is unchanged as the
asymptotic value of the scalar is varied.

Let us end with two remarks.  In the AdS case one can hope that there
is a dual description for the attractor phenomenon.  Since the
asymptotic value of the scalar is changing we are turning on a
operator in the dual theory with a varying value for the coupling
constant. The fact that the entropy, for fixed charge, does not change
means that the number of ground states in the resulting family of dual
theories is the same.  This would be worth understanding in the dual
description better.  Finally, we expect this analysis to generalise in
a straightforward manner to the AdS space in higher dimensions as
well.

Fig. 6 illustrates the attractor mechanism in asymptotically $AdS_4$
space.  This Figure is for the example, eq.(\ref{se}, \ref{seffpot}).
The cosmological constant is taken to be, $\Lambda =-2.91723$, in
$\kappa=1$ units.

\begin{figure}[htb]
  \begin{center}
    \epsfig{file=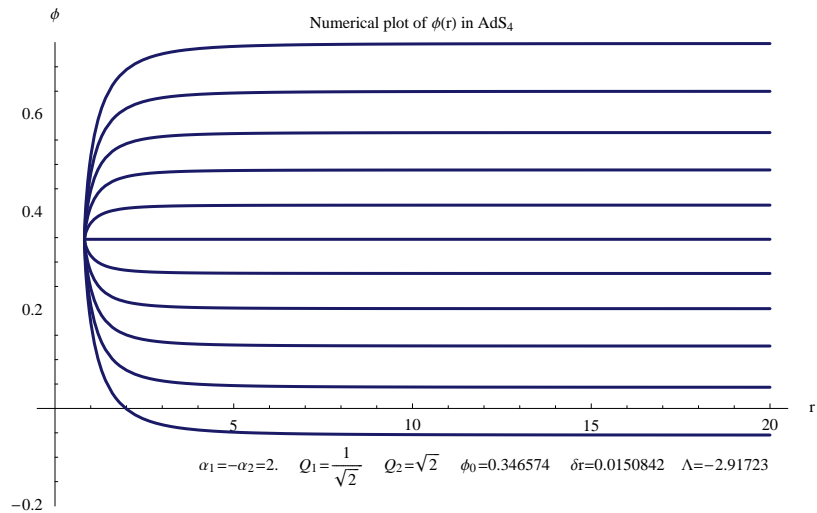,width=0.8\linewidth}
  \end{center}
  \caption{\label{cap:ads}Numerical plot of $\phi(r)$ with
    $\alpha_1=-\alpha_2=2$ for the an extremal black hole in $AdS_4$
    displaying attractor behaviour. }
\end{figure}

\section{Additional Comments}
The theories we considered in the discussion of asymptotically flat
space-times and AdS spacetimes have no potential for the scalars.  We
comment on this further here.

Let us consider a theory with ${\cal N}=1$ supersymmetry containing
chiral superfields whose lowest component scalars are,
\begin{equation}
  \label{defcs}
  S_i=\phi_i+i a_i
\end{equation}
We take these scalars to be uncharged under the gauge symmetries.
These can be coupled to the superfields $W_\alpha^a$ by a coupling
\begin{equation}
  \label{gcoup}
  L_{gauge kinetic}=\int d^2\theta f_{ab}(S_i) W_\alpha^a W_\alpha^b
\end{equation}
Such a coupling reproduces the gauge kinetic energy terms in and
eq.(\ref{fourdact}), eq.(\ref{actoned}), (we now include both $\phi_i,
a_i$ in the set of scalar fields which we denoted by $\phi_i$ in the
previous sections).

An additional potential for the scalars would arise due to F-term
contributions from a superpotential.  If the superpotential is absent
we get the required feature of no potential for these scalar.  Setting
the superpotential to be zero is at least technically natural due to
its non-renormalisability.

In a theory with no supersymmetry there is no natural way to suppress
a potential for the scalars and it would arise due to quantum effects
even if it is absent at tree-level.  In this case we have no good
argument for not including a potential for the scalar and our analysis
is more in the nature of a mathematical investigation.

The absence of a potential is important also for avoiding no-hair
theorems which often forbid any scalar fields from being excited in
black hole backgrounds \cite{Nohair}. In the presence of a mass $m$ in
asymptotically flat four dimensional space the two solutions for first
order perturbation at asymptotic infinity go like,
\begin{equation}
  \label{assmass}
  \phi \sim C_1 e^{mr}/r, \phi \sim C_2 e^{-mr}/r.
\end{equation}
We see that one of the solutions blows up as $r \rightarrow \infty$.
Since one solution to the equation of motion also blows up in the
vicinity of the horizon, as discussed in section 2, there will
generically be no non-singular solution in first order perturbation
theory. This argument is a simple-minded way of understanding the
absence of scalar hair for extremal black holes under discussion here.
In the absence of mass terms, as was discussed in section 2, the two
solutions at asymptotic infinity go like $ \phi \sim {\rm const}$ and
$\phi \sim 1/r$ respectively and are both acceptable.  This is why one
can turn on scalar hair. The possibility of scalar hair for a massless scalar 
is of course well known. See \cite{Gibbons:1985ac}, \cite{Gibbons:1987ps},
 for some early examples of solutions with scalar hair, \cite{Masood-ul-Alam:1993ea, Mars:2001pz, 
Gibbons:2002ju,Gibbons:2002av},
 for theorems on uniqueness in the presence of such hair, and \cite{Gibbons:1996af}  
for a discussion of 
resulting thermodynamics.  

In asymptotic AdS space the analysis is different. Now the $(mass)^2$
for scalars can be negative as long as it is bigger than the BF bound.
In this case both solutions at asymptotic infinity decay and are
acceptable. Thus, as for the massless case, it should be possible to
turn on scalar fields even in the presence of these mass terms and
study the resulting black holes solutions.  Unfortunately, the
resulting equations are quite intractable. For small $(mass)^2$ we
expect the attractor mechanism to continue to work.
 
If the $(mass)^2$ is positive one of the solutions in the asymptotic
region blows up and the situation is analogous to the case of a
massive scalar in flat space discussed above.  In this case one could
work with AdS space which is cut off at large $r$ (in the infrared)
and study the attractor phenomenon.  Alternatively, after
incorporating back reaction, one might get a non-singular geometry
which departs from AdS in the IR and then analyse black holes in this
resulting geometry.  In the dual field theory a positive $(mass)^2$
corresponds to an irrelevant operator.  The growing mode in the bulk
is the non-normalisable one and corresponds to turning on a operator
in the dual theory which grows in the UV. Cutting off AdS space means
working with a cut-off effective theory. Incorporating the
back-reaction means finding a UV completion of the cut-off theory.
And the attractor mechanism means that the number of ground states at
fixed charge is the same regardless of the value of the coupling
constant for this operator.

\section{Asymptotic de Sitter Space}
In de Sitter space the simplest way to obtain a double-zero horizon is
to take a Schwarzschild black hole and adjust the mass so that the de
Sitter horizon and the Schwarzschild horizon coincide. The resulting
black hole is the extreme Schwarzschild-de Sitter spacetime
\cite{exds}.  We will analyse the attractor behaviour of this black
hole below.  The analysis simplifies in 5-dimensions and we will
consider that case, a similar analysis can be carried out in other
dimensions as well.  Since no charges are needed we set all the gauge
fields to zero and work only with a theory of gravity and scalars.  Of
course by turning on gauge charges one can get other double-zero
horizon black holes in dS, their analysis is left for the future.

We start with the action of the form,
\begin{equation}
  \label{actds}
  S=\frac{1}{\kappa^{2}}\int d^{5}x\sqrt{-G}(R-2(\partial\phi)^{2}-V(\phi))
\end{equation}
Notice that the action now includes a potential for the scalar,
$V(\phi)$, it will play the role of $V_{eff}$ in our discussion of
asymptotic flat space and AdS space.  The required conditions for an
attractor in the dS case will be stated in terms of $V$. A concrete
example of a potential meeting the required conditions will be given
at the end of the section.  For simplicity we have taken only one
scalar, the analysis is easily extended for additional scalars.

The first condition on $V$ is that it has a critical point,
$V'(\phi_0)=0$.  We will also require that $V(\phi_0) >0$.  Now if the
asymptotic value of the scalar is equal to its critical value,
$\phi_0$, we can consistently set it to this value for all times $t$.
The resulting equations have a extremal black hole solution mentioned
above. This takes the form
\begin{equation}
  \label{nariamet}
  ds^2=-{t^2 \over (t^2/L-L/2)^2 } dt^2 +{(t^2/L-L/2)^2 \over t^2} dr^2 + t^2 d\Omega_3^2
\end{equation}
Notice that it is explicitly time dependent.  $L$ is a length related
to $V(\phi_0)$ by , $V(\phi_0)={20 \over L^2}$. And $t=\pm {L \over
  \sqrt{2}}$ is the location of the double-zero horizon. A suitable
near-horizon limit of this geometry is called the Nariai solution,
\cite{nariai}.

\subsection{Perturbation Theory}
Starting from this solution we vary the asymptotic value of the
scalar. We take the boundary at $t \rightarrow -\infty$ as the initial
data slice and investigate what happens when the scalar takes a value
different from $\phi_0$ as $t \rightarrow -\infty$. Our discussion
will involve part of the space-time, covered by the coordinates in
eq.(\ref{nariamet}), with $-\infty\le t \le t_H= -{L \over \sqrt{2}}$.
We carry out the analysis in perturbation theory below.

Define the first order perturbation for the scalar by, $$
\phi=\phi_0
+ \epsilon \phi_1$$
This satisfies the equation,
\begin{equation}
  \label{fods}
  \partial_t(a_0^2b_0^3\partial_t\phi_1)={b^3\over 4} V''(\phi_0)\phi_1
\end{equation}
where $a_0={(t^2/L-L/2)\over t}$, $b_0=t$.  This equation is difficult
to solve in general.

In the vicinity of the horizon $t=t_H$, we have two solutions which go
like,
\begin{equation}
  \label{twosolds}
  \phi_1=C_{\pm} (t-t_H)^{-1+\sqrt{1+\kappa^2} \over 2}
\end{equation}
where
\begin{equation}
  \label{valkds}
  \kappa^2=-{1\over 4} V''(\phi_0)
\end{equation}
We see that one of the two solutions in eq.(\ref{twosolds}) is
non-divergent and in fact vanishes at the horizon if
\begin{equation}
  \label{condpotds}
  V''(\phi_0) <0.
\end{equation}
We will henceforth assume that the potential meets this condition.
Notice this condition has a sign opposite to what was obtained for the
asymptotically flat or AdS cases.  This reversal of sign is due to the
exchange of space and time in the dS case.

In the vicinity of $t \rightarrow -\infty$ there are two solutions to
eq.(\ref{fods}) which go like,
\begin{equation}
  \label{foinds}
  \phi_1={\tilde C}_{\pm}|t|^{p_{\pm}}
\end{equation}
where
\begin{equation}
  \label{valpm}
  p_{\pm}=2(-1 \pm \sqrt{1+\kappa^2/4}).
\end{equation}
If the potential meets the condition, eq.(\ref{condpotds}) then
$\kappa^2>0$ and we see that one of the modes blows up at $t
\rightarrow -\infty$.

\subsection{Some Speculative  Remarks}
In view of the diverging mode at large $|t|$ one needs to work with a
cutoff version of dS space \footnote{This is related to some comments
  made in the previous section in the positive $(mass)^2$ case in AdS
  space.}.  With such a cutoff at large negative $t$ we see that there
is a one parameter family of solutions in which the scalar takes a
fixed value at the horizon. The one parameter family is obtained by
starting with the appropriate linear combination of the two solutions
at $t \rightarrow -\infty$ which match to the well behaved solution in
the vicinity of the horizon.  While we will not discuss the metric
perturbations and scalar perturbations at second order these too have
a non-singular solution which preserves the double-zero nature of the
horizon.  The metric perturbations also grow at the boundary in
response to the growing scalar mode and again the cut-off is necessary
to regulate this growth. This suggests that in the cut-off version of
dS space one has an attractor phenomenon. Whether such a cut-off makes
physical sense and can be implemented appropriately are question we
will not explore further here.

One intriguing possibility is that quantum effects implement such a
cut-off and cure the infra-red divergence. The condition on the
potential eq.(\ref{condpotds}) means that the scalar has a negative
$(mass)^2$ and is tachyonic. In dS space we know that a tachyonic
scalar can have its behaviour drastically altered due to quantum
effects if it has a $(mass)^2<H^2$ where $H$ is the Hubble scale of dS
space. This can certainly be arranged consistent with the other
conditions on the potential as we will see below. In this case the
tachyon can be prevented from ``falling down'' at large $|t|$ due to
quantum effects and the infrared divergences can be arrested by the
finite temperature fluctuations of dS space.  It is unclear though if
any version of of the attractor phenomenon survives once these quantum
effects became important.

We end by discussing one example of a potential which meets the
various conditions imposed above.  Consider a potential for the
scalar,
\begin{equation}
  \label{potds}
  V=\Lambda_1e^{\alpha_1 \phi} + \Lambda_2 e^{\alpha_2 \phi}.
\end{equation}
We require that it has a critical point at $\phi=\phi_0$ and that the
value of the potential at the critical point is positive. The critical
point for the potential eq.(\ref{potds}) is at,
\begin{equation}
  \label{critds}
  e^{\phi_{0}}=-\left( {\alpha_2 \Lambda_2 \over \alpha_1 \Lambda_1}\right)^{1\over \alpha_1-\alpha_2}
\end{equation} 
Requiring that $V(\phi_0)>0$ tells us that
\begin{equation}
  \label{tcds}
  V(\phi_0)
  =\Lambda_2 e^{\alpha_2 \phi_0}\left(1-{\alpha_2 \over \alpha_1}\right) >0
\end{equation}
Finally we need that $V''(\phi_0) <0$ this leads to the condition,
\begin{equation}
  \label{sdds}
  V''(\phi_0)=\Lambda_2e^{\alpha_2\phi_0}\alpha_2(\alpha_2-\alpha_1)<0
\end{equation}
These conditions can all be met by taking both $\alpha_1, \alpha_2>0$,
$\alpha_2<\alpha_1$, $\Lambda_2>0$ and $\Lambda_1<0$.  In addition if
$\alpha_2\alpha_1 \gg 1$ the resulting $-(mass)^2 \gg H^2$.

\section{Non-Extremal = Unattractive }\label{sec:nonex}

We end the paper by examining the case of an non-extremal black hole
which has a single-zero horizon.  As we will see there is no attractor
mechanism in this case. Thus the existence of a double-zero horizon is
crucial for the attractor mechanism to work.

Our starting point is the four dimensional theory considered in
section 2 with action eq.(\ref{actiongen}).  For simplicity we
consider only one scalar field.  We again start by consistently
setting this scalar equal to its critical value, $\phi_0$, for all
values of $r$, but now do not consider the extremal Reissner Nordstrom
black hole.  Instead we consider the non-extremal black hole which
also solves the resulting equations.  This is given by a metric of the
form, eq.(\ref{metric2}), with
\begin{equation}
  \label{metne}
  a^2(r)=\left(1-{r_+\over r}\right)\left({1-{r_-\over r}}\right),\qquad b(r)=r
\end{equation}
where $r_\pm$ are not equal. We take $r_+>r_-$ so that $r_+$ is the
outer horizon which will be of interest to us.
 
The first order perturbation of the scalar field satisfies the
equation,
\begin{equation}
  \label{fone}
  \partial_r(a^2b^2\partial_r\phi_i)={V_{eff}''(\phi_0)\over 4 b^2} \phi_1
\end{equation}
In the vicinity of the horizon $r=r_+$ this takes the form,
\begin{equation}
  \label{vhne}
  \partial_y(y \partial_y \phi_1)=\alpha \phi_1
\end{equation}
where $\alpha$ is a constant dependent on $V''(\phi_0), r_+,r_-$, and
$y \equiv r-r_+$.

This equation has one non-singular solution which goes like,
\begin{equation}
  \label{nhsne}
  \phi_1 = C_0 + C_1 y + \cdots
\end{equation}
where the ellipses indicate higher terms in the power series expansion
of $\phi_1$ around $y=0$.  The coefficients $C_1, C_2 , \cdots$ are
all determined in terms of $C_0$ which can take any value.  Thus we
see that unlike the case of the double-horizon extremal black hole,
here the solution which is well-behaved in the vicinity of the horizon
does not vanish.

Asymptotically, as $r \rightarrow \infty$ both solutions to
eq.(\ref{fone}) are well defined and go like $1/r, constant$
respectively. It is then straightforward to see that one can choose an
appropriate linear combination of the two solutions at infinity and
match to the solution, eq.(\ref{nhsne}) in the vicinity of the
horizon. The important difference here is that the value of the
constant $C_0$ in eq.(\ref{nhsne}) depends on the asymptotic values of
the scalar at infinity and therefore the value of $\phi$ does not go
to a fixed value at the horizon. The metric perturbations sourced by
the scalar perturbation can also be analysed and are non-singular. In
summary, we find a family of non-singular black hole solutions for
which the scalar field takes varying values at infinity.  The crucial
difference is that here the scalar takes a value at the horizon which
depends on its value at asymptotic infinity.  The entropy and mass for
these solutions also depends on the asymptotic value of the scalar
\footnote{An intuitive argument was given in the introduction in
  support of the attractor mechanism.  Namely, that the degeneracy of
  states cannot vary continuously. This argument only applies to the
  ground states. A non-extremal black hole corresponds to excited
  states.  Changing the asymptotic values of the scalars also changes
  the total mass and hence the entropy in this case.}.

It is also worth examining this issue in a non-extremal black holes
for an exactly solvable case.

If we consider the case $|\alpha_i|=2$, section \ref{sec:exact_sol},
the non-extremal solution takes on a relatively simple form.  It can
be written\cite{Kallosh:1992ii}
\begin{eqnarray}
  \label{eq:nonex}
  \exp(2\phi) & = & e^{2\phi_{\infty}}\frac{(r+\Sigma)}{(r-\Sigma)}\nonumber \\
  a^{2} & = & \frac{(r-r_{+})(r-r_{-})}{(r^{2}-\Sigma^{2})}\label{eq:alpha2_solution}\\
  b^{2} & = & (r^{2}-\Sigma^{2})\nonumber 
\end{eqnarray}
where\footnote{The radial coordinate $r$ in
  eq.(\ref{eq:alpha2_solution}) is related to our previous one by a
  constant shift.}
\begin{equation}
  \label{rhne}
  r_{\pm}=M\pm r_{0}\qquad r_{0}=\sqrt{M^{2}+\Sigma^{2}-\bar{Q}_2^{2}-\bar{Q}_1^{2}}.
\end{equation}
and the Hamiltonian constraint becomes
\begin{equation}
  \Sigma^{2}+M^{2}-\bar{Q}_{1}^{2}-\bar Q_{2}^{2}=\frac{1}{4}(r_+-r_-)^{2}.
\end{equation}
The scalar charge, $\Sigma$, defined by
$\phi\sim\phi_\infty+\frac{\Sigma}{r} $, is not an independent
parameter.  It is given by
\begin{equation}
  \Sigma=\frac{\bar{Q}_2^{2}-\bar{Q}_1^{2}}{2M}.
\end{equation} 
There are horizons at $r=r_{\pm}$, the curvature singularity occurs at
$r=\Sigma$ and $r_{0}$ characterises the deviation from extremality.
We see that the non-extremal solution does not display attractor
behaviour.

\begin{figure}[htb]
  \begin{center}
    \epsfig{file=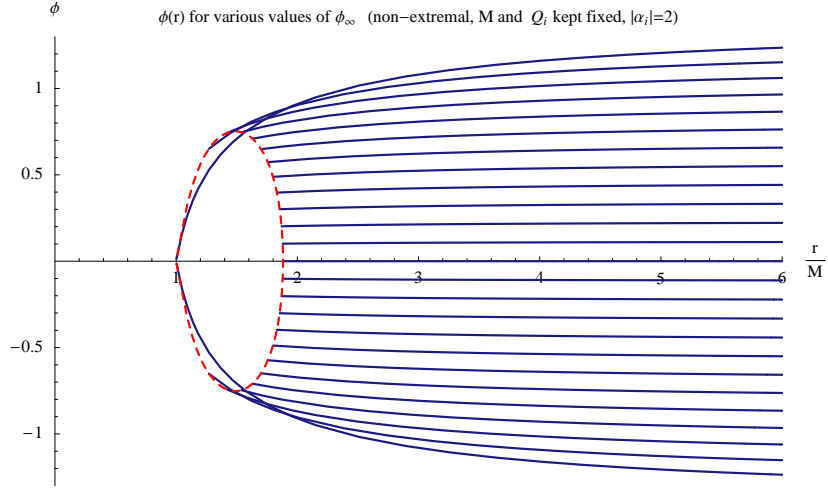,width=0.8\linewidth}
  \end{center}
  \caption{\label{cap:unattractive}Plot $\phi(r)$ with
    $\alpha_1=-\alpha_2=2$ for the non-extremal black hole with $M,
    Q_i$ held fixed while varying $\phi_\infty$.  The dotted line
    denotes the outer horizon at which we terminate the plot.  It is
    clearly unattractive.}
\end{figure}

Fig. \ref{cap:unattractive} shows the behaviour of the scalar field
\footnote{for $\alpha_1=-\alpha_2=2$} as we vary $\phi_\infty$ keeping
$M$ and $Q_i$ fixed. The location of the horizon as a function of $r$
depends on $\phi_\infty$, eq.(\ref{rhne}).  The horizon as a function
of $\phi_\infty$ is denoted by the dotted line. The plot is terminated
at the horizon.

In contrast, for the extremal black hole,
\begin{equation}
  \label{eq:alpha2:Mass}
  M=\frac{|\bar{Q}_2|+|\bar{Q}_1|}{\sqrt{2}}\qquad\Sigma=\frac{|\bar Q_2|-|\bar Q_1|}{\sqrt{2}},
\end{equation}
so (\ref{eq:nonex}) gives
\begin{equation}
  \label{eq:phi:oncemore}
  e^{2\phi_0}=e^{2\phi_\infty}\frac{M+\Sigma}{M-\Sigma}
  \stackrel{(\ref{eq:alpha2:Mass})}{=}
  \frac{|Q_2|}{|Q_1|},
\end{equation}
which is indeed the attractor value.

\bigskip \goodbreak \centerline{\bf Acknowledgements}
\noindent
We thank A.  Dabholkar, R. Gopakumar, G. Gibbons, S. Minwalla, A. Sen,
M. Shigemori and A.  Strominger for discussions.  N.I. carried out
part of this work while visiting Harvard University.  He would like to
thank the Harvard string theory group for its hospitality and
support.  We thank the organisers of ISM-04, held at Khajuraho, for a
stimulating meeting. This research is supported by the Government of India. 
S.T. acknowledges support from the Swarnajayanti Fellowship, DST, Govt. of India. 
Most of all we thank the people of India for
generously supporting research in String Theory.

\appendix
\section{Perturbation Analysis}\label{sec:appa}
\renewcommand{\theequation}{A.\arabic{equation}}
\setcounter{equation}{0}

\subsection{Mass}
\label{Mass_calc}

Here, we first calculate the mass of the extremal black hole discussed
in section 2.2.  From eq.(\ref{finalb2}), for large r, $b_{2}$ is
given by,
\begin{equation}
  b_{2}=cr+d\label{assb2}
\end{equation}
where
\begin{eqnarray}
  c & = & -\frac{c_{1}^{2}\gamma}{2(2\gamma-1)}\\
  d & = & \frac{r_H c_{1}^{2}\gamma^2}{(2\gamma-1)}
\end{eqnarray}
Now, we can easily write down the expression for $a_{2}$ using
eq.(\ref{a2}).  We choose coordinate $y$ as introduced in
eq.(\ref{by}) such that at large r,
\begin{eqnarray}
  r^{2}+2\epsilon^{2}(cr^{2}+dr) & = & y^{2}\\
  \frac{1}{r} & = & \frac{1}{y}(1+\epsilon^{2}(c+\frac{d}{y}))\end{eqnarray}
We use the extremality condition (\ref{finala}) to find,
\begin{eqnarray}
  a(r)=\left(\frac{r-r_H}{y}\label{assa2}\right)\end{eqnarray}

Using, eq.(\ref{assb2}, \ref{assa2}) one finds that asymptotically, as
$r\rightarrow \infty$ the metric takes the form,
\begin{equation}
  \label{assmetr}
  ds^2=-\left(1-{2(r_H+\epsilon^2 (c r_H+d)) \over y} \right) d{\tilde t}^2 + 
  {1 \over (1-{2(r_H+\epsilon^2 (c r_H+d)) \over y} )} dy^2 + y^2 d\Omega^2
\end{equation} 
where ${\tilde t}$ is obtained by rescaling $t$ and $d\Omega^2$
denotes the metric of $S^2$.  The mass $M$ of the black hole is then
given by the $1/y$ term in the $g_{yy}$ component of the metric.  This
gives,
\begin{equation}
  M=r_H+\epsilon^{2}\frac{r_Hc_1^2\gamma}{2}. \end{equation}

\subsection{Perturbation Series to All Orders}
\label{sec:all_orders}

Next we go on to discuss the perturbation series to all orders, Using
(\ref{pertb}) for $b$ and (\ref{pertphi}) for $\phi$ in
eq.(\ref{eq2})and eq.(\ref{eqphi}), we get,
\begin{eqnarray}
  b_{k}''=-\sum_{i=0}^{k}\sum_{j=0}^{k-i}b_{i}\phi_{j}'\phi_{k-i-j}'
\end{eqnarray}
\begin{eqnarray}
  \sum_{i+j\leqslant k }
  2b_{j}b_{k-i-j}((r-r_H)^{2}\phi_{i}')'
  = Q_i^2 e^{\alpha_i\phi_0}\alpha_i  {\cal V}_{ik}
\end{eqnarray}
where
\begin{eqnarray}
  \label{eq:def:s_k}
  {\cal V}_{ik} &=& \sum_{\stackrel{\{n_1, n_2 \ldots n_{k}\}}{\sum m n_m = k}}
  \frac{\phi_1^{n_1}\phi_2^{n_2}\ldots \phi_k^{n_{k}}}{n_1!n_2!\ldots n_{k}!}\alpha^{n_1+n_2+\ldots+n_{k}}_i.\\
\end{eqnarray}
After substituting our ansatz (\ref{diln})and (\ref{bn}), the above
equations give,
\begin{equation}
  \label{eq:dk_rel}
  k (k \gamma-1)d_k = -\gamma\sum_{{i+j< k} } j(k-i-j)d_i c_j c_{k-i-j}  
\end{equation}
and
\begin{equation}
  \label{eq:aftersubs}
  k(k\gamma+1)c_k+
  T_k= (\gamma+1)(c_k+ S_k)
\end{equation}
where $S_k$ and $T_k$ are given by
\begin{eqnarray}
  \label{eq:def:Sk}
  S_k &=&
  \sum_{\stackrel{\{n_1, n_2 \ldots n_{k-1}\}}{\sum m n_m = k}}
  \frac{c_1^{n_1}c_2^{n_2}\ldots c_{k-1}^{n_{k-1}}}{n_1!n_2!\ldots n_{k-1}!}
  \left(
    \alpha^{\sum n_l-1}_1 +\alpha^{\sum n_l-1}_2
  \right)
\end{eqnarray}
and
\begin{equation}
  \label{eq:def:Tk}
  T_k=\sum_{\stackrel{j+l\leqslant k}{l<k}}l(l\gamma+1)d_j d_{k-l-j}c_l.
\end{equation}
Then solving for $d_k$ and $c_k$ gives
\begin{eqnarray}
  d_{k}&=&-\frac{ \gamma}{k(k\gamma-1)}\sum_{{i+j< k} } j(k-i-j)d_i e_j e_{k-i-j}
  \label{dk_ans} \\  
  c_k &=&\frac{(\gamma+1)S_k-T_k}{((k+1)\gamma+1)(k-1)}
  \label{ek_ans}
\end{eqnarray}
Finally, $e_k$ can be obtained using eq.(\ref{finala}), eq.(\ref{bn}).
It can be verified that the ansatz, eq.(\ref{diln}, \ref{bn},
\ref{an}) with the coefficients eq.(\ref{dk_ans}, \ref{ek_ans}) also
solves the constraint eq.(\ref{constraint}).


\section{Exact Analysis }\label{sec:details}
\renewcommand{\theequation}{B.\arabic{equation}}
\setcounter{equation}{0}

Exact solutions can be found by writing the equations of motion as
generalised Toda equations \cite{Toda}, which may, in certain special
cases, be solved exactly \cite{Gibbons:1987ps} - we rederive this
result in slightly different notation below. As noted in
\cite{Lu:1996jr}, in a marginally different context, the extremal
solutions, are, in appropriate variables, polynomial solutions of the
Toda equations.  The polynomial solutions are much easier to find and
are related to the functions $f_i$ mentioned in section
\ref{sec:exact_sol}. For ease of comparison we occasionally use
notation similar to \cite{Lu:1996jr}.

\subsection{New Variables}

To recast the equations of motion into a generalised Toda equation we
define the following new variables
\begin{equation}
  u_1=\phi
  \qquad u_2=\log a
  \qquad z=\log ab
  \qquad\cdot=\partial_{\tau}=a^{2}b^{2}\partial_{r}
\end{equation}
In terms of $r$, $\tau$ is given by
\begin{equation}
  \tau=\int\frac{dr}{a^{2}b^{2}}=\frac{1}{(r_{+}-r_{-})}\log\left(\frac{r-r_{+}}{r-r_{-}}\right)
\end{equation}
where $r_\pm$ are the integration constants of (\ref{eq1}). In general
(\ref{eq1}) implies
\begin{equation}
  \label{eq:def:rpm}
  a^2 b^2 = (r-r_+)(r-r_-).
\end{equation}
Notice that
\begin{eqnarray}
  \tau\rightarrow0 & \mbox{as} & r\rightarrow\infty\\
  \tau\rightarrow-\infty & \mbox{as} & r\rightarrow r_{+}
\end{eqnarray}
When we have a double-zero horizon, $r_{H}=r_{\pm}$, $\tau$ takes the
simple form
\begin{equation}
  \label{eq:tau_double}
  \tau^{-1}=-{\left(r-r_{H}\right)}.
\end{equation}
Since we are mainly interested in solutions with double-zero horizons,
in what follows it will be convenient to work with a new radial
coordinate, $\rho$, defined by
\begin{equation}
  \label{eq:def:rho}
  \rho=-\tau^{-1}.
\end{equation}
which has the convenient property that $\rho_H=0$.

\subsection{Equivalent Toda System}
In terms of these new variables the equations of motion become
\begin{eqnarray}
  \ddot{u}_1 &=&
  \textha\alpha_1 e^{2u_2+\alpha_1 u_1}Q_{1}^{2}+\textha\alpha_2e^{2 u_2 +\alpha_2u_1}Q_{2}^{2} \label{eq:phi_eom_3}\\
  \ddot{u}_2 &=& e^{2u_2+\alpha_1 u_1}Q_{1}^{2}+e^{2u_2+\alpha_2u_1}Q_{2}^{2} \label{eq:v_eom_1}\\
  \ddot{z}&=&e^{2z}  \label{eq:z_EOM}
\end{eqnarray}
\begin{equation}
  \dot{u_1}^{2}+\dot{u_2}^{2}-\dot{z}^{2}
  +e^{2z}-e^{2u_2+\alpha_1 u_1}Q_{1}^{2}-e^{2u_2+\alpha_2 u_1}Q_{2}^{2}=0
  \label{eq:toda_E}
\end{equation}
(\ref{eq:z_EOM}) decouples from the other equations and is equivalent
to (\ref{eq1}).  Finally making the coordinate change
\begin{equation}
  \label{eq:def:X}
  X_i = n_{ij}^{-1}u_j +m_{ij}^{-1}\log\left((\alpha_1-\alpha_2)Q_{j}^{2}\right)
\end{equation}
where
\begin{equation}
  \label{eq:def:n}
  n^{-1}=\left(
    \begin{array}{cc}
      2 & -\alpha_2 \\
      -2 & \alpha_1
    \end{array}\right)
\end{equation}
and
\begin{equation}
  m_{ij}=\frac{1}{2\left(\alpha_1-\alpha_2\right)}
  \left(
    4+\alpha_i\alpha_j
  \right)
\end{equation}
we obtain the generalised 2 body Toda equation
\begin{equation}
  \ddot{X}_{i}=e^{m_{ij}X_{j}},
\end{equation}
together with
\begin{equation}
  \label{eq:toda_ham}
  \sum_{ij}\left(\textstyle\ha \dot X_i m_{ij}\dot X_j 
    -e^{m_{ij} X_j}\right)=(\alpha_1-\alpha_2)\mathcal{E}
\end{equation}
where ${\cal E}=\frac{1}{4}(r_+-r_-)^2$.
%
After solving the above, the original fields will be given by
\begin{eqnarray}
  \label{eq:final_sol}
  e^{(\alpha_1-\alpha_2)\phi}&=&\frac{Q_2^2}{Q_1^2}e^{\ha(\alpha_1 X_1 +\alpha_2 X_2)} \\
  a^2 &=& e^{\frac{2}{\alpha_1-\alpha_2}(X_1+X_2)}/\diamondsuit \\
  b^2 &\stackrel{(\ref{eq1})}{=}& (r-r_+)(r-r_-)/a^2
\end{eqnarray}
where
\begin{equation}
  \label{eq:defdiamond}
  \diamondsuit = (\alpha_1-\alpha_2)Q_1^{2\foob} Q_2^{2\fooa}
\end{equation}

\subsection{Solutions}

\subsubsection{Case I: $\gamma=1\Leftrightarrow \alpha_1\alpha_2=-4$}

In this case, $m_{ij}$ is diagonal
\begin{equation}
  \label{eq:M_gamma1}
  m=\mathrm{diag}(\alpha_1/2,-\alpha_2/2),
\end{equation}
so the equations of motion decouple:
\begin{equation}
  \label{eq:eom:gamma1}
  \ddot{X}_{i}=e^{\frac{|\alpha_i|}{2} X_{i}}.
\end{equation}
(\ref{eq:eom:gamma1}) has solutions
\begin{equation}
  X_i = \frac{2}{|\alpha_i|}
  \log\left(
    \frac{4c_i^2}{|\alpha_i|\sinh^2(c_i(\tau-d_i))}
  \right)
\end{equation}
The integration constants are fixed by imposing asymptotic boundary
conditions and requiring that the solution is finite at the horizon.
Letting
\begin{equation}
  F_{i}=\sinh(c_{i}\left(\tau-d_{i}\right))
\end{equation}
in terms of $\phi$ and $a$ we get
\begin{equation}
  \label{eq:exact_sol_case1}
  \begin{array}{cll}
    e^{(\alpha_1-\alpha_2)\phi}&=&\displaystyle\frac{Q_2^2}{Q_1^2}e^{\ha(\alpha_1 X_1 +\alpha_2 X_2)} \\
    &=& \displaystyle
    \left(
      -\frac{\alpha_2}{\alpha_1}
    \right) 
    \left(
      \frac{Q_2 F_2 c_1}{Q_1 F_1 c_2}
    \right)^2 \\
    a^2 &=&\displaystyle\left.e^{\frac{2}{\alpha_1-\alpha_2}(X_1+X_2)}\right/ \diamondsuit\\
    &=& 
    \displaystyle
    \left.
      \left(\frac{ c_1}{Q_1 F_1}\right)^{2\foob} 
      \left(
        \frac{ c_2}{Q_2 F_2}
      \right)^{2\aod}
    \right/\spadesuit
  \end{array}
\end{equation}
As $r\rightarrow r_{+}$(ie. $\tau\rightarrow-\infty$) the scalar field
goes like
\begin{equation}
  e^{(\alpha_1-\alpha_2)\phi}  \sim  e^{2(c_{1}-c_{2})\tau}\\
\end{equation}
so we require
\begin{equation}
  c:= c_{1}  =  c_{2}
\end{equation}
for a finite solution at the horizon. Also at the horizon
\begin{equation}
  \label{eq:b_at_horiz}
  b^{2} \sim (r-r_+)/a^2 \sim  e^{\left((r_{+}-r_{-})-2c\right)\tau}  
\end{equation}
which necessitates
\begin{equation}
  (r_{+}-r_{-})  =  2c
\end{equation}
To find the extremal solutions we take the limit $c\rightarrow0$ which
gives
\begin{eqnarray}
  e^{(\alpha_1-\alpha_2)\phi} 
  & = & \left(
    -\frac{\alpha_2}{\alpha_1}\right)\frac{\left(Q_{2}f_{2}\right)^{2}}{\left(Q_{1}f_{1}
    \right)^{2}}\\
  b^{2} 
  & = & \spadesuit \left(
    Q_{1}f_{1}\right)^{\frac{-2{\alpha_2}}{\alpha_1-\alpha_2}}\left(Q_{2}p_{2}
  \right)^{\frac{2\alpha_1}{\alpha_1-\alpha_2}}\\
  a^{2} & = & \rho^{2}/b^{2}
\end{eqnarray}
where
\begin{equation}
  \label{eq:fi}
  f_{i}  =  1+d_{i}\rho.
\end{equation}
Requiring $\phi\rightarrow\phi_\infty$ and $a\rightarrow1$ as
$r\rightarrow\infty$ fixes
\begin{equation}
  d_{i}=\bar{Q_{i}}^{-1}\sqrt{\frac{|\alpha_{i}|}{\alpha_1-\alpha_2}}
\end{equation}
where as before
\begin{equation}
  \bar{Q}_{i}^2=e^{\alpha_{i}\phi_{\infty}}Q_{i}^2\nosum.
\end{equation}

For comparison with the non-extremal solution in this case see section
\ref{sec:nonex}.

\subsubsection{Case II: $\gamma=2$ and $\alpha_1=-\alpha_2=2\sqrt{3}$
}
In this case, $m_{ij}$ becomes
\begin{equation}
  \label{eq:M_gamma2}
  m=\left(
    \matrix{ \frac{2}
      {{\sqrt{3}}} & -
      \frac{1}{{\sqrt{3}}}
      \cr -
      \frac{1}{{\sqrt{3}}}
      & \frac{2}
      {{\sqrt{3}}} \cr  }
  \right).
\end{equation}
It is convenient to use the coordinates
\begin{equation}
  \label{eq:def:q_gamma2}
  q_i = \frac{1}{\sqrt 3}X_i - \sqrt3\log\sqrt3 
\end{equation}
so the equations of motion are the two particle Toda equations
\begin{eqnarray}
  \label{eq:eom:gamma2}
  \ddot{q}_{1}&=&e^{2 q_{1}-q_2}\\
  \ddot{q}_{2}&=&e^{2 q_{2}-q_1}.
\end{eqnarray}
These maybe integrated exactly but the explicit form is, in general, a
little complicated. Fortunately we are mainly interested in extremal
solutions which have a simpler form \cite{Lu:1996jr}. As in,
\cite{Lu:1996jr}, taking the ansatz that $e^{-q_i}$ is a second order
polynomial one finds
\begin{eqnarray}
  \label{eq:gamma2_anzatz}
  e^{-q_1}&=&a_0+a_1\tau+\ha\tau^2 \\
  e^{-q_2}&=&a_1^2-a_0+a_1\tau+\ha\tau^2
\end{eqnarray}
Finally, returning to the original variables and imposing the
asymptotic boundary conditions gives the solution
\begin{eqnarray}
  e^{4\sqrt{3}\phi} & = & \left(\frac{\PP}{\Q}\right)^2
  \left(\frac{f_2}{f_1}\right)^{6}\label{eq:exact_scalar_sol_gamma2}\\
  b^{2} & = & 2Q_1 Q_2 f_1 f_2\label{eq:exact_b_sol_gamma2}\\
  a^{2} &=& \rho^2/b^2
\end{eqnarray}
where
\begin{equation}
  f_i  = \left( 1+
    (\bar{Q}_1 \bar{Q}_2)^{-\frac{2}{3}} (\bar{Q}_1^{\frac{2}{3}} + \bar{Q_2}^{\frac{2}{3}})^\ha \rho
    +{\textstyle\ha} (\bar{Q}_i\bar{Q}_1 \bar{Q}_2)^{-\frac{2}{3}}\rho^{2}\right)^\ha
\end{equation} 
as quoted in section \ref{sec:exact_sol}.

For completeness we note that the general, non-extremal solution of
\cite{Dobiasch:1981vh,Gibbons:1985ac}, modified for a non-zero
asymptotic value of $\phi$, is
\begin{eqnarray}
  \exp(4\phi/\sqrt{3}) & =e^{4\phi_{\infty}/\sqrt{3}} & \frac{p_2}{p_1}\\
  a^{2} & = & \frac{(r-r_{+})(r-r_{-})}{\sqrt{p_1 p_2}}\\
  b^{2} & = & \sqrt{p_1 p_2}
\end{eqnarray}
where
\begin{equation}
  p_i=(r-r_{i+})(r-r_{i-})
\end{equation}
\begin{equation}
  r_{i\pm}=\frac{2}{(-\alpha_i)}\Sigma\pm\bar{Q}_i\sqrt{\frac{4\Sigma}{2\Sigma+\alpha_i M}}
\end{equation}
and scalar charge, $\Sigma$, which is again not an independent
parameter, is given by
\begin{equation}
  \frac{1}{\sqrt{3}}\Sigma=\frac{\bar{Q}_2^{2}}{2M(\lambda-1)}+\frac{\bar{Q}_1^{2}}{2M(\lambda+1)}\qquad\lambda=\frac{\Sigma}{\sqrt{3}M}
\end{equation}

\subsubsection{Case III: $\gamma=3$ and $\alpha_1=4\qquad\alpha_2=-6$
}
In this case, $m_{ij}$ becomes
\begin{equation}
  m=\left(
    \begin{array}{cc}
      1 & -1\\
      -1 & 2
    \end{array}\right)
\end{equation}
Making the coordinate change
\begin{eqnarray}
  \label{eq:def:q_gamma3}
  q_1 &=& \textha X_1 - \log 2 \\
  q_2 &=& X_2 - \log 2
\end{eqnarray}
The equations of motion are
\begin{eqnarray}
  \label{eq:gamma3_eom}
  \ddot q_1 &=& e^{2q_1-q_2} \\
  \ddot q_2 &=& e^{2q_2-2q_2} 
\end{eqnarray}
Now consider the three particle Toda system
\begin{eqnarray}
  \ddot q_1 &=& e^{2q_1-q_2} \label{eq:toda_gamma3_1}\\
  \ddot q_2 &=& e^{2q_2-q_1-q_3} \\
  \ddot q_3 &=& e^{2q_3-q_2} \label{eq:toda_gamma3_2}
\end{eqnarray} 
which may be integrated exactly. Notice that by identifying $q_1$ and
$q_3$ we obtain (\ref{eq:toda_gamma3_1}-\ref{eq:toda_gamma3_2}). Once
again the general solution is slightly complicated but taking the
ansatz that $e^{-q_i}$ is a polynomial one finds
\begin{eqnarray}
  e^{-q_{1}} & = & a_{0}+2a_{2}^{2}\tau+a_{2}\tau^{2}+\frac{1}{6}\tau^{3}\\
  e^{-q_{2}} 
  & = & 4a_{2}^{4}-2a_{0}a_{2}
  +(4a_{2}^{3}-a_{0})\tau+2a_{2}^{2}\tau^{2}+\frac{2}{3}a_{2}\tau^{3}+\frac{1}{12}\tau^{4}
\end{eqnarray}

Rewriting in terms of the original fields we get
\begin{eqnarray}
  e^{10\phi} & = & \left(\frac{Q_{2}}{Q_{1}}\right)^{2}\exp\left(2X_{1}-3X_{2}\right)\\
  & = & \frac{6}{4}\left(\frac{Q_{2}}{Q_{1}}\right)^{2}\left(\frac{f_{2}}{f_{1}}\right)^{12}\\
  b^{2} 
  & = & \rho^2 10 Q_{1}^{\frac{6}{5}}Q_{2}^{\frac{4}{5}}
  \exp\left(-\frac{1}{5}X_{1}-\frac{1}{5}X_{2}\right)\\
  &= &\frac{5}{2}
  \left(\frac{2}{3}\right)^{\frac{3}{5}}Q_{1}^{\frac{6}{5}}Q_{2}^{\frac{4}{5}}f_1 f_2 \\
  a^2 &=& \rho^2/b^2
\end{eqnarray} 
where
\begin{equation}
  f_{1} =  \left(
    1-6a_{2}\rho + 12a_{2}^{2}\rho^{2}-6a_{0}\rho^{3}
  \right)^{\frac{1}{3}}
\end{equation}
\begin{equation}
  f_{2}  =  \left(
    1-{\textstyle \frac{24}{3}}a_{2}\rho
    +24a_{2}\rho^{2}-(48a_{2}^{3}-12a_{0})\rho^{3}+\left(48a_{2}^{4}-24a_{0}a_{2}\right)\rho^{4}
  \right)^{\frac{1}{4}}
\end{equation}
At the horizon we do indeed have $\phi$ at the critical point of
$V_{eff}$:
\begin{equation}
  e^{10\phi_0}=\frac{3}{2}\frac{Q_{2}^{2}}{Q_{1}^{2}}
\end{equation}
and $b^2$ given by $V_{eff}(\phi_0)$:
\begin{equation}
  b_H^{2}=
  \frac{5}{2}\left(\frac{2}{3}\right)^{\frac{3}{5}}
  Q_{1}Q_{2}\left(\frac{Q_{2}}{Q_{1}}\right)^{\frac{1}{5}}.
\end{equation}
Imposing the asymptotic boundary conditions we get
%
\begin{equation}
  a_{0}=\pm\frac{2^{\frac{5}{7}}}{\bar{Q}_{1}^{\frac{10}{7}}\bar{Q}_{2}^{\frac{5}{7}}}
  \qquad
  \left(4a_{2}^{4}-2a_{0}a_{2}\right)
  =\frac{2^{\frac{11}{7}}}{\bar{Q}_{1}^{\frac{22}{7}}\bar{Q}_{2}^{\frac{18}{7}}}
\end{equation}
so letting
\begin{eqnarray}
  \clubsuit & = & \frac{2^{\frac{11}{7}}}{Q_{1}^{\frac{22}{7}}Q_{2}^{\frac{18}{7}}}\\
  \Delta_{1} & = & 3\sqrt[3]{a_{0}^{3}+\sqrt{a_{0}^{6}+\frac{64}{3}a_{0}^{3}\clubsuit^{3}}}\\
  \Delta_{2} & = & \sqrt{\frac{3^{\frac{1}{3}}\Delta_{1}}{a_0}-\frac{3^{\frac{2}{3}}4}{\Delta_1}}
\end{eqnarray}
we may write $a_2$ as
\begin{equation}
  a_{2}
  =\pm\frac{1}{2\sqrt{6}}\Delta_{2}\pm\frac{1}{2}
  \sqrt{
    \frac{2\clubsuit}{3^{\frac{1}{3}}\Delta_{1}}
    -\frac{\Delta_{1}}{2\,3^{\frac{2}{3}}}+\frac{\sqrt{6}}{\Delta_{2}}
  }
\end{equation}
Despite the non-trivial form of the solution we see that it still
takes on the attractor value at the horizon.

In terms of the $U(1)$ charges (written implicitly in terms of $a_0$
and $a_2$), the mass and scalar charge are expressed below


\begin{equation}
  \Sigma=\frac{3a_{0}^{2}-28a_{0}a_{2}^{3}+32a_{2}^{6}}{40a_{0}a_{2}^{4}-20a_{0}^{2}a_{2}}
\end{equation}


\begin{equation}
  M=
  \frac{
    (a_{0}^{2}+4a_{0}a_{2}^{3}-16a_{2}^{6})
  }
  {
    2^{\frac{3}{5}}50a_{0}^{\frac{7}{5}}a_{2}(a_{0}-2a_{2}^{3})
    (2a_{2}^{4}-a_{0}a_{2})^{\frac{1}{5}}
    Q_{1}^{\frac{6}{5}}Q_{2}^{\frac{4}{5}}
  }
\end{equation}

This solution is related to a 3 charge p-brane solution found in
\cite{Lu:1996jr} - in this case we have identified two of the degrees
of freedom.
\section{Higher Dimensions }\label{sec:higherdim}
\renewcommand{\theequation}{C.\arabic{equation}}
\setcounter{equation}{0}
Here we give some more details related to our discussion of the higher
dimensional attractor in section 5.  The Ricci components calculated
from the metric, eq.(\ref{metrichd}) are,
\begin{eqnarray}
  R_{tt} & = & a^{2}\left(a'^{2}+{\frac{{(d-2)aa'b'}}{{b}}}+aa''\right) \\
  R_{rr} & = & -\{ b\left(a'^{2}+aa''\right)+(d-2)a\left(a'b'+ab''\right)\}/a^{2}b  \\
  R_{\theta\theta} & = &
  (d-3)-2aba'b'-a^{2}\left((d-3)b'^{2}+bb''\right)
\end{eqnarray}

The Einstein equations from the action eq.(\ref{higherdaction}), take
the form,
\begin{eqnarray}
  R_{tt} & = & \frac{(d-3)(d-3)!a(r)^{2}}{b(r)^{2(d-2)}}V_{eff}(\phi_i) \\
  R_{rr} & = & 2(\partial_{r}\phi)^{2}-\frac{(d-3)(d-3)!}{b(r)^{2(d-2)}a(r)^{2}}V_{eff}(\phi_i) \nonumber \\
  \\
  R_{\theta\theta} & = & \frac{(d-3)!}{b^{2(d-3)}}V_{eff}(\phi_i),
\end{eqnarray}
where $V_{eff}$ is given by eq.(\ref{eact}).

Taking the combination, ${1\over 2}(R_{rr}-{G_{rr}\over
  G_{tt}}R_{tt})$ gives, eq.(\ref{Rrreq}).  Similarly we have,
\begin{eqnarray}
  \label{Hconstraint}
  &  & {\frac{b(r)^{2}}{a(r)^{2}}}R_{tt}+a(r)^{2}b(r)^{2}R_{rr}-(d-2)R_{\theta\theta}\nonumber \\
  & = & -(d-2)\{ d-3-a(r)b'(r)(2a'(r)b(r)+(d-3)a(r)b'(r))\}\nonumber \\
  & = & 2(\partial_{r}\phi_i)^{2}a(r)^{2}b(r)^{2}-\frac{(d-2)(d-3)!}{b^{2(d-3)}}V_{eff}(\phi_i)\
\end{eqnarray}
This gives eq.(\ref{constrainthd}).  Finally the relation,
$R_{tt}=(d-3){a^2\over b^2}R_{\theta\theta}$ yields,
\begin{eqnarray}
  &  & (d-3)^{2}(-1+a(r)^{2}b'(r)^{2})+b(r)^{2}(a'(r)^{2}+a(r)a''(r))\nonumber \\
  &  & +a(r)b(r)((-8+3d)a'(r)b'(r)+(d-3)a(r)b''(r))=0. \label{leqhd}\end{eqnarray}
 
We now discuss solving for $a_2$, the second order perturbation in the
metric component $a$, in some more detail. We restrict ourselves to
the case of one scalar field, $\phi$.  The constraint,
eq.(\ref{constrainthd}), to $O(\epsilon^2)$ is,
\begin{eqnarray}
  (d-2) ra_{2}'+(d-2)(d-3)a_{2}-2(\phi_{1}')^{2}r^{2}(1-(\frac{r_{H}}{r})^{d-3})^{2}\label{constrainthigher1storder}\\
  -2(d-2)(d-3)^{2}\frac{r_{H}^{2(d-3)}}{r^{2(d-3)+1}}b_{2}+2(d-3)^{2}\frac{\gamma(\gamma+1)\phi_{1}^{2}}{r^{2(d-3)}r_{H}^{6-2d}}\nonumber \\
  +2(d-2)\frac{(d-3)(r_{H}^{3}r^{d}-r_{h}^{d}r^{3})}{r_{H}^{6}r^{2d}}\left\{ r_{H}^{d}r^{2}b_{2}+r_{h}^{3}r^{d}b_{2}'\right\}  =  0\nonumber \end{eqnarray}
This is a first order equation for $a_2$ of the form,
\begin{equation}
  f_{1}a_{2}'+f_{2}a_{2}+f_{3}=0,
\end{equation}
where,
\begin{eqnarray}
  f_{1} & = & (d-2) r \nonumber \\
  f_{2} & = & (d-2)(d-3)\nonumber \\
  f_{3} & = & -2(\phi_{1}')^{2}r^{2}(1-(\frac{r_{H}}{r})^{d-3})^{2}-2(d-2)(d-3)^{2}\frac{r_{H}^{2(d-3)}}{r^{2(d-3)+1}}b_{2}\nonumber \\
  &  & +2(d-3)^{2}\frac{t(t+1)\phi_{1}^{2}}{r^{2(d-3)}r_{H}^{6-2d}} \nonumber \\
  & & +2(d-2)\frac{(d-3)(r_{H}^{3}r^{d}-r_{H}^{d}r^{3})}{r_{H}^{6}r^{2d}}\left\{ r_{H}^{d}r^{2}b_{2}+r_{H}^{3}r^{d}b_{2}'\right\} 
  \label{f1f2f3}\end{eqnarray}

The solution to this equation is given by,
\begin{equation}
  \label{sola2hd}
  a_{2}(r)=Ce^{\mathcal{F}}-e^{\mathcal{F}}\int e^{-\mathcal{F}}\frac{f_{3}}{f_{1}}dr
\end{equation}
where $\mathcal{F}=-\int\frac{f_{2}}{f_{1}}dr$.  It is helpful to note
that $e^{\mathcal{F}}= {1\over r^{(d-3)}}$ and,
$\frac{e^{-\mathcal{F}}}{f_{1}}={r^{d-4} \over (d-2)}$.
 
Now the first term in eq.(\ref{sola2hd}), proportional to $C$, blows
up at the horizon. We will omit some details but it is easy to see
that the second term in eq.(\ref{sola2hd}) goes to zero. Thus for a
non-singular solution we must set $C=0$.  One can then extract the
leading behaviour near the horizon of $a_2$ from eq.(\ref{sola2hd}),
however it is slightly more convenient to use eq.(\ref{leqhd}) for
this purpose instead. From the behaviour of the scalar perturbation
$\phi_1$, and metric perturbation, $b_2$, in the vicinity of the
horizon, as discussed in the section on attractors in higher
dimensions, it is easy to see that
\begin{equation}
  a_{2}(r)=A_{2}(r^{d-3}-r_{H}^{d-3})^{2\gamma+2}
\end{equation}
where, $A_2$ is an appropriately determined constant.
Thus we see that the non-singular solution in the vicinity of the
horizon vanishes like $(r-r_H)^{(2\gamma+2)}$ and the double-zero
nature of the horizon persists after including back-reaction to this
order.

Finally, expanding eq.(\ref{sola2hd}) near $r\rightarrow \infty$ (with
$C=0$) we get that $a_2 \rightarrow {\rm Const} +
{\mathcal{O}}(1/r^{d-3})$.  The value of the constant term is related
to the coefficient in the linear term for $b_2$ at large $r$ in a
manner consistent with asymptotic flatness.

In summary we have established here that the metric perturbation $a_2$
vanishes fast enough at the horizon so that the black hole continues
to have a double-zero horizon, and it goes to a constant at infinity
so that the black hole continues to be asymptotically flat.

\section{More Details on Asymptotic AdS Space} \label{sec:ads}
\renewcommand{\theequation}{D.\arabic{equation}}
\setcounter{equation}{0}
We begin by considering the asymptotic behaviour at large $r$ of
$\phi_1$, eq.(\ref{seads}).  One can show that this is given by
\begin{equation}
  \label{asads}
  \phi_{1}(r)\rightarrow 
  c_{+}{\frac{1}{r^{3/2}}}I_{3/4}\left(\frac{\kkappa L}{2r^{2}}\right)
  +c_{-}{\frac{1}{r^{3/2}}}I_{-3/4}\left(\frac{\kkappa L}{2r^{2}}\right)
\end{equation}

Here $I_{3/4}$ stands for a modified Bessel function
\footnote{Modified Bessel function $I_{\nu}(r),K_{\nu}(r)$ does
  satisfy following differential eq.
  \begin{equation}
    z^{2}I_{\nu}''(z)+zI_{\nu}'(z)-(z^{2}+\nu^{2})I(z)=0.
  \end{equation}}
Asymptotically, $I_\nu \propto r^{-2\nu}$. Thus $\phi_r$ has  two solutions 
which go asymptotically to a constant and as $1/r^3$ respectively. 

Next, we consider values of r, $r_H<r<\infty$.  These are all ordinary
points of the differential equation eq.(\ref{seads}).  Thus the
solution we are interested is well-behaved at these points.  For a
differential equation of the form,
\begin{equation}
  \label{deord}
  \mathcal{L}(\psi)=\frac{d^{2}\psi}{dz^{2}}+p(z)\frac{d\psi}{dz}+q(z)\psi=0,
\end{equation}
all values of $z$ where $p(z),q(z)$ are analytic are ordinary points.
About any ordinary point the solutions to the equation can be expanded
in a power series, with a radius of convergence determined by the
nearest singular point \cite{MF}.
 
We turn now to discussing the solution for $a_2$.  The constraint
eq.(\ref{constraintads}) takes the form,
\begin{equation}
  2a_{0}^{2}b_{2}'+a_{2}+ (a_0^2)'(r b_2)'+ra_{2}'
  =\frac{-1}{r^{2}}\kkappa^{2}\phi_{1}^{2}+a_{0}^{2}
  r^{2}(\partial_{r}\phi_{1})^{2}+{2 b_2 \over r^3}(r_H^2+{2r_H^4 \over L^2}) +\frac{6rb_{2}}{L^2}
\end{equation}

The solution to this equation is given by,
\begin{equation}
  \label{sola2ads}
  a_{2}(r)=\frac{c_{2}}{r}-\frac{1}{r}\int_{r_H} f_{3}dr
\end{equation}
where
\begin{equation}
  f_{3}=2a_{0}^{2}b_{2}'+ (a_0^2)'(rb_2)'
  +\frac{1}{r^{2}}\beta^{2}\phi_{1}^{2}-a_{0}^{2}r^{2}(\partial_{r}\phi_{1})^{2}
  -{2 b_2 \over r^3}(r_H^2+{2 r_H^4 \over L^2})-\frac{6rb_{2}}{L^2}
\end{equation}.
We have set the lower limit of integration in the second term at
$r_H$.  We want a solution the preserves the double-zero structure of
the horizon. This means $c_2$ must be set to zero.

To find an explicit form for $a_2$ in the near horizon region it is
slightly simpler to use the equation, eq.(\ref{eq1ads}).  In the near
horizon region this can easily be solved and we find the solution,
\begin{equation}
  \label{asha2ads}
  a_2 \propto (r-r_H)^{(2 \gamma + 2)}.
\end{equation}

At asymptotic infinity one can use the integral expression,
eq.(\ref{sola2ads}) (with $c_2=0$).  One finds that $f_3 \rightarrow r
$ as $r \rightarrow \infty$.  Thus $a_2 \rightarrow d_2 r$.  This is
consistent with the asymptotically AdS geometry.

In summary we see that that there is an attractor solution to the
metric equations at second order in which the double-zero nature of
the horizon and the asymptotically AdS nature of the geometry both
persist.


\end{document}